\documentclass[]{article}
\usepackage[a4paper, total={6in, 8in}]{geometry}
\usepackage[version=4]{mhchem}
\usepackage[]{hyperref}
\usepackage{xcolor}
\usepackage{soul}
\usepackage[normalem]{ulem}
\hypersetup{
    colorlinks,
    linkcolor={red!50!black},
    citecolor={blue!50!black},
    urlcolor={blue!80!black}
}
\usepackage[htt]{hyphenat} 
\usepackage{graphicx} 
\usepackage{amsmath}
\usepackage{mathtools}
\graphicspath{{Figures/}}
\usepackage{wrapfig}
\usepackage{comment}
\usepackage{float}
\usepackage{enumitem}
\usepackage{amsfonts}
\usepackage[sort&compress,comma,numbers,super]{natbib}
\usepackage{tabularx}
\usepackage{authblk}  

\title{Leveraging Large Language Models to Address Data Scarcity in Machine Learning: Applications in Graphene Synthesis}
\author[1]{Devi Dutta Biswajeet}
\author[1]{Sara Kadkhodaei\thanks{Corresponding author: \texttt{sarakad@uic.com}}}

\affil[1]{Department of Civil, Materials, and Environmental Engineering, University of Illinois Chicago, Chicago, IL, United States of America}

\date{}

\begin{document}
\maketitle

\section*{Abstract}
Machine learning in materials science faces challenges due to limited experimental data, as generating synthesis data is costly and time-consuming, especially with in-house experiments. Mining data from existing literature offers a potential solution but introduces issues like mixed data quality, inconsistent formats, and variations in reporting experimental parameters, complicating the creation of consistent features for the learning algorithm. Additionally, combining continuous and discrete features can hinder the learning process with limited data. Here, we propose strategies that utilize large language models (LLMs) to enhance machine learning performance on a limited, heterogeneous dataset of graphene chemical vapor deposition synthesis compiled from existing literature. These strategies include prompting modalities for imputing missing data points and leveraging large language model embeddings to encode the complex nomenclature of substrates reported in chemical vapor deposition experiments. Imputed data from various prompting modalities are compared to statistical K-nearest neighbors imputation, with the former yielding a more diverse distribution, richer feature representation, and improved model generalization. We also demonstrate that discretizing continuous features enhances prediction accuracy. The proposed strategies enhance graphene layer classification using a support vector machine (SVM) model, increasing binary classification accuracy from 39\% to 65\% and ternary accuracy from 52\% to 72\%. We compare the performance of the SVM and a GPT-4 model, both trained and fine-tuned on the same data. Our results demonstrate that the numerical classifier, when combined with LLM-driven data enhancements, outperforms the standalone LLM predictor, highlighting that improving predictive learning with LLM strategies requires more than simple fine-tuning on datasets in data-scarce scenarios. Instead, it necessitates sophisticated approaches for data imputation and feature space homogenization to achieve optimal performance. The proposed strategies emphasize data enhancement techniques rather than merely refining learning architectures or regularizing loss functions, offering a broadly applicable framework for improving machine learning performance on scarce, inhomogeneous datasets.
\\
\textbf{Keywords:} Chemical Vapor Deposition; Graphene; Data Scarcity in Machine Learning; Large Language Models; Data Imputation; Text Embeddings

\section{Introduction}\label{sec:introduction}

Machine learning on scarcely populated datasets often leads to deterioration of learning performance, including poor generalization and high bias. 
The predictive accuracy of a model often goes down with the reduction of training data size, and the errors increase following a power law \cite{zhang2018strategy}. Models with high variance and strong learning capacity often struggle with overfitting on small datasets, achieving near-perfect accuracy on the training data but failing to generalize effectively to unseen data. 
In the materials machine learning literature, strategies for addressing data scarcity are typically implemented at three levels\cite{NANDY2022100778,Xu2023,Chen2024}: the data level, such as data augmentation and synthetic data generation\cite{Oviedo2019,Ohno2020,Ma2020,Davariashtiyani2023}, manual or automatic data extraction from publications\cite{Zeng2015,Swain2016,Matic2016,Tao2021,Kononova2021,Hong2021}, and crowd-sourcing data repositories\cite{schiller2020crowd}; the algorithm level, such as loss function or algorithm regularization\cite{Chen2021}, feature engineering\cite{Zhang2018,DeBreuck2021,Gong2023,Anand2022}, Bayesian approaches\cite{olivier2021bayesian, zhou2022towards} including both aleatoric and epistemic uncertainties into the prediction framework, and dimensionality reduction\cite{Rajan2009}; and the machine learning strategy level\cite{hutchinson2017,pmlr-v119-tsai20a}, such as $\Delta$ machine learning\cite{Liu2023,Grumet2024}, transfer learning\cite{Gupta2021,Gong2022,Pan2023,Magar2022,Feng2021}, active and reinforced learning \cite{Wang2024,Chai2024}, and mixture-of-experts\cite{Chang2022,Cheenady2024}. 

Large language models (LLMs) have recently emerged as a new approach in predictive chemistry and materials science \cite{Zheng2023,D3DD00202K,Lee2024,Polak2024,Kim2024,Jablonka2024,Antunes2024,Kang2024,gruver2024}. Specifically, they hold significant potential to address data scarcity and heterogeneity challenges, particularly by utilizing the extensive embedded pre-trained knowledge to impute missing data or homogenize inconsistencies in reported features. 
LLMs are referred to as near-universal generalists \cite{yu2024large}, positioning them as readily accessible alternatives to the aforementioned strategies in materials informatics. Furthermore, LLMs can be fine-tuned for specific problems with minimal prompt engineering, making them technically feasible for navigating complex feature spaces \cite{jablonka202314}. This flexibility and generality offer potential solutions to both underfitting and overfitting challenges in machine learning on small datasets while also enabling LLMs to minimize hallucinations during data generation or featurization. In essence, LLMs could help address the bias-variance tradeoff that most ML strategies struggle to resolve when working with limited data.
In this study, we demonstrate an LLM-assisted enhancement of machine learning using a sparsely populated and inherently heterogeneous dataset on chemical vapor deposition (CVD) of graphene, compiled from multiple studies in the literature. 


The CVD synthesis of graphene involves a wide range of experimental setups, with several key parameters governing the growth of graphene layers \cite{dobkin2003principles,BHOWMIK2022103832,zhang2013review}. To this date, no comprehensive dataset encompassing an exhaustive set of synthesis parameters has been developed for graphene CVD growth or synthesizing any 2D (atomic-layer-thick) material. Consequently, most machine learning studies on 2D material growth rely on limited, in-house experimental data, often constrained to specific ranges of synthesis conditions (e.g., pressures, temperatures, and precursor flow rates). These studies typically utilize machine learning models that require minimal feature engineering, as the in-house experimental data is already labeled and well-structured \cite{lu2022machine, xu2021machine, tang2020machine, wu2024universal,hwang2021machine}. For example, Hwang \textit{et al.} developed a machine learning model to predict the morphology of graphene in the form of SEM (Scanning Electron Microscopy) images based on CVD processing variables (i.e., temperature, annealing time, growth time, and supply of hydrogen) \cite{hwang2021machine}. The size, coverage, domain density, and aspect ratio of CVD-grown graphene in SEM images were automatically measured using a region proposal convolutional neural network (R-CNN). A neural network and support vector machine were trained to map the relationship between CVD processing variables and these morphological features. Subsequently, a generative adversarial network (GAN) was designed to generate SEM images based on specified CVD processing conditions \cite{hwang2021machine}. However, this study relied on in-house experimental data, restricted to four processing variables and limited by fixed chamber conditions, including a constant precursor flow rate and the exclusive use of a single substrate type (Cu). Another notable effort by Schiller \textit{et al.} involved creating a crowd-sourced database compiling an extensive list of parameters influencing various aspects of graphene CVD growth, contributed by multiple experimental groups \cite{schiller2020crowd}. They also developed a UNet model for SEM image characterization, accessible as a nanoHub tool \cite{schiller_gresq}. However, the database remains incomplete, making it challenging to integrate it into a machine-learning framework for predicting outcomes based on CVD processing conditions.

Machine learning approaches must be tailored to handle limited data effectively in the absence of a large dataset with comprehensive experimental parameters for graphene CVD synthesis. In this study, we demonstrate how large language models can be leveraged to enhance the performance of a classification task on limited data. We compiled a database encompassing varied conditions of pressure, temperature, substrates, and precursor flow rates. Tabular datasets from several experimental studies were identified \cite{saeed2020chemical, hong2023recent, son2017low, liu2011synthesis, zhang2013review, liu2017controlled, bhuyan2016synthesis, chen2015large, huang2021substrate}, and relevant entries were manually extracted through careful data mining, resulting in a heterogeneous and small dataset, as detailed in section \ref{sec:method}. The dataset offers sufficient feature variability (inputs) to serve as a starting point for training a machine learning model. However, it is hindered by the inhomogeneity of the mined data, leading to numerous missing values across nearly all attributes except for the substrate. To address this challenge, data imputation techniques are explored using large language models and statistical methods such as K-nearest neighbors (as detailed in section \ref{sec:method}). As detailed in section \ref{sec:results}, the LLM-based approach offers greater flexibility for data imputation, resulting in a more diverse dataset. In contrast, the KNN method merely replicates the underlying data distribution, producing a dataset with limited variability that remains constrained by the scarcity of the original data.

Recent advancements in large language models enable them to perform complex tasks such as understanding context and semantics and generating contextually relevant responses. The vast training data, pretraining on diverse corpora, and the sheer scale of these models make them excel at interpreting meaning and delivering accurate, coherent outputs. 
Several studies have utilized LLMs for imputation tasks \cite{nazir2023chatgpt,hayat2024claim, jacobsen2024imputation, fang2025spatiotemporal, ding2024data, wang2025llm, chen2023gatgpt}, showing superior performance compared to traditional techniques such as K nearest neighbor (KNN), mean imputation, multivariate imputation by chained equations (MICE) and ML methods such as generative and discriminative methods (GAN) \cite{he2024llm, wang2025llm, hayat2024claim, ji2024predicting}.
For example, Nazir \textit{et al.} utilized \texttt{ChatGPT-3.5} as an imputation engine to populate missing values for biological and psychological data, showcasing its out-performance compared to KNN and mean imputation\cite{nazir2023chatgpt}. 
They applied an iterative human-in-the-loop feedback process to refine \texttt{ChatGPT}'s imputation response by comparing the preceding response with ground-truth data, resulting in context-aware and accurate imputations of missing values in subsequent steps. 
Another study by Hayat \textit{et al.}\cite{hayat2024claim} introduced an LLM-based imputation formalism to populate missing values for seven multivariate classification datasets from the UCI repository\cite{kelly2024uci}. 
They generated missing-value descriptors using a pre-trained LLaMA2 model and then fine-tuned the model using the descriptor-enriched data set for classification tasks. The classification tasks were directly performed using the fine-tuned LLMs. 
Other notable studies have incorporated techniques such as QnA prompts for data imputation and property extraction using LLMs across a wide variety of tasks \cite{ji2024predicting, sipilä2024questionansweringmodelsinformation, he2024llm}.  
This study focuses on leveraging the existing pre-trained \texttt{ChatGPT-4o-mini} for a general imputation strategy. Additionally, the learning capacities of numerical models, such as support vector machines and random forests, are evaluated for the target classification task and are compared with a fine-tuned \texttt{ChatGPT-4o-mini} counterpart. 


Another challenge in our small dataset, beyond data imputation, is the inconsistent representation of attributes like the substrate. This inconsistency arises because the data is sourced from multiple studies with varying nomenclatures for substrate reporting. To address this, we leverage LLMs to uniformly featurize the substrate. These textual attributes are processed using embedding models, which convert them into meaningful vector representations that serve as features. By incorporating OpenAI’s embedding models, we achieve improved classification accuracy compared to traditional label encoding methods like one-hot encoding, which can inflate feature dimensionality and lead to overfitting, particularly in small datasets. As a final step, the continuous input feature space is transformed into a discrete map, a process demonstrated to enhance the classification performance of the machine learning model.

\begin{figure}[h!]
    \centering
    \includegraphics[width=\textwidth]{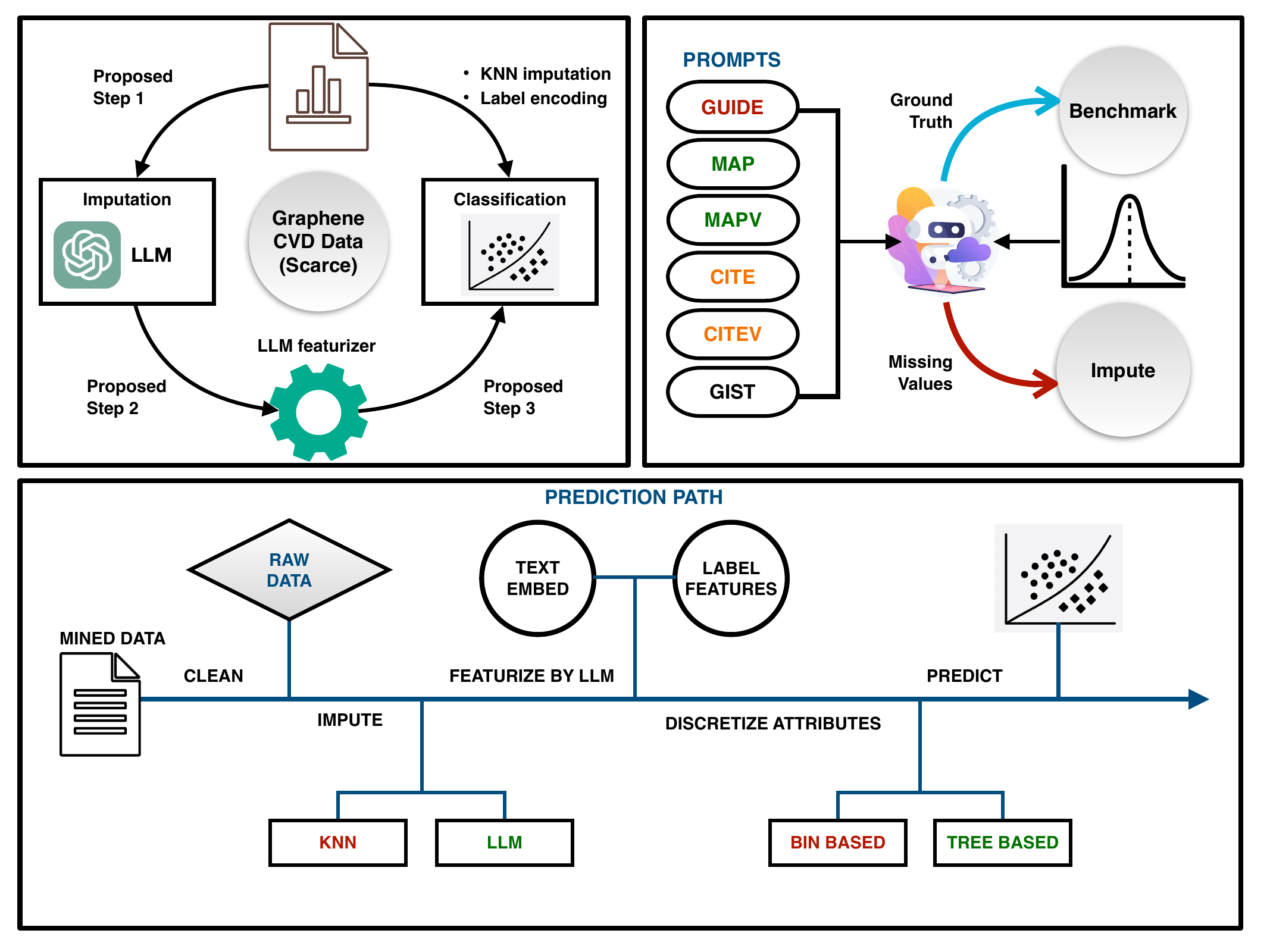}  
    \caption{Overview of the methods employed in this study to enhance predictive classification performance on a limited, heterogeneous dataset for graphene chemical vapor deposition growth.}
    \label{fig:overview}  
\end{figure}

As illustrated in Figure \ref{fig:overview}, the workflow of this study involves several key steps.
A small data set for CVD graphene synthesis is manually curated from the literature through meticulous data mining. 
LLM-based data imputation is benchmarked against KNN-based imputation for the classification task. For the LLM-based approach, various prompt-engineering strategies guide the imputation task with specific instructions. After imputation, LLM-based featurization is used to unify and homogenize the numerical features for training binary and ternary classifiers, which predict the number of graphene layers based on a set of CVD processing inputs. We compare the effectiveness of different prompt-engineering strategies with traditional KNN-based data imputation, highlighting the superior performance of specific prompt-engineered GPT models. The prompt-driven imputed values are benchmarked with existing ground truth, and the missing values are directly imputed as an independent task. LLM-based strategies minimize imputation errors while enhancing data diversity, yielding a knowledge-driven, less imitative data distribution that outperforms KNN in machine learning. Additionally, we compare traditional non-LLM label-encoded features, commonly used for handling inconsistent labels, with LLM-generated features. Our results demonstrate the superior performance of LLM-embedded features in enhancing ML performance compared to label- or category-encoded features. Finally, we discretize the attributes to enhance the model's learning performance compared to their continuous counterparts. These techniques improve ML performance for graphene layer classification, as detailed in the following sections. 
%

\section{Methodology}\label{sec:method}
\subsection{Data Collection}
Experimental literature was used to manually extract 164 data entries \cite{saeed2020chemical, hong2023recent, son2017low, liu2011synthesis, zhang2013review, liu2017controlled, bhuyan2016synthesis, chen2015large, huang2021substrate}, each consisting of 10 attributes. Five of these attributes represent precursor flow rates (sccm), while the remaining chamber parameters include pressure (mbar), temperature ($^{\circ}\text{C}$), growth time (min), and substrate. The target variable is the number of graphene layers. Many studies also report the CVD method; however, this primarily reflects the pressure range used in the CVD process. Therefore, we did not include it as a separate attribute. Table \ref{tab:graphene_growth_data} shows a subset of the dataset used in this study. The dataset contains missing values for pressure, growth time, number of layers, and nearly all precursor flow rates. The precursor flow rates for each data entry are presented in the  \texttt{graphene\_growth\_conditions\_layers.csv}
supplementary file. The raw data distributions across various attributes are shown in Supplementary Figure 1. More details on data collection are provided in Supplementary Note 1.
\begin{table}[H]
\centering
\resizebox{\textwidth}{!}{%
\begin{tabular}{|c|c|c|c|c|c|}
\hline
\textbf{CVD Method} & \textbf{Pressure (mbar)} & \textbf{Temperature ($^{\circ}$C)} & \textbf{Growth Time (min)} & \textbf{Substrate} & \textbf{No. of Layers} \\
\hline
Hot-Wall & 800.0 & 766.67 &  & Ge/Si wafer & ML \\
\hline
Hot-Wall & 750.0 & 1050.0 & 30.0 & Cu & ML \\
\hline
Hot-Wall & 1013.25 & 1070.0 & 60.0 & Ni & 3.5 \\
\hline
Hot-Wall & 26.67 & 1000.0 &  & Cu & ML \\
\hline
Hot-Wall & 13.33 & 1000.0 & 4.0 & Cu & ML \\
\hline
Hot-Wall & 133.3 & 975.0 & 1.5 & Cu & 1 \\
\hline
Hot-Wall & 53.32 & 1030.0 & 30.0 & Cu & ML \\
\hline
Hot-Wall &  & 1060.0 & 60.0 & Cu & 1 \\
\hline
Hot-Wall & 13.33 & 1020.0 &  & Cu & ML \\
\hline
Hot-Wall & 1013.25 & 1080.0 & 5.0 & Cu & 1 \\
\hline
Hot-Wall &  & 1040.0 & 67.5 & In, Ga; Cu & 1 \\
\hline
Hot-Wall & 1013.25 & 1020.0 & 35.0 & Ga & 1 \\
\hline
Hot-Wall &  & 1100.0 & 30.0 & Cu &  \\
\hline
Hot-Wall &  & 1062.5 & 380.0 & Pt3Si/Pt & 2 \\
\hline
Hot-Wall &  & 1080.0 & 15.0 & Cu & ML \\
\hline
\vdots &\vdots &\vdots & \vdots&\vdots &\vdots \\
\hline
\end{tabular}%
}
\caption{A subset of the graphene CVD growth dataset compiled from various literature sources \cite{saeed2020chemical, hong2023recent, son2017low, liu2011synthesis, zhang2013review, liu2017controlled, bhuyan2016synthesis, chen2015large, huang2021substrate}.}
\label{tab:graphene_growth_data}
\end{table}

\subsection{Imputation of Missing Attributes}
Missing data for each attribute was imputed using one statistical method (KNN) and six prompt-based LLM methods. KNN algorithm identifies the most similar data points, or neighbors, to a given data point with missing values and estimates the missing feature using the mean or median of the corresponding features from its $K$-nearest neighbors. 
The $K$-nearest neighbors are identified using the Euclidean distance $d$ between data points $i$ and $j$, which is calculated as:
\begin{equation}
    d(i, j) = \sqrt{\sum_{k=1}^n (x_{ik} - x_{jk})^2}
\end{equation}
where $x_{ik}$ and  $x_{jk}$ are  the values of feature $k$ for data points $i$ and $j$, respectively, among $n$ total features. The missing value is then estimated by the mean or median of feature $k$ from its K-nearest neighbors:
\begin{equation}
    x_{i, k - \text{missing}} = \frac{1}{K} \sum_{j \in N(i)} x_{j,k}
\end{equation}
where $x_{j, k - \text{missing}}$ represents the values of feature $k$ from the neighbors of data point $i$, denoted as $N(i)$, that have non-missing values. The choice of $K$ involves a bias-variance trade-off, where lower values of $K$ result in higher variance, while higher values of $K$ introduce greater bias. Therefore, $K=5$ was chosen as a rule of thumb for our dataset containing 164 entries. The choice of $K$ can vary depending on the complexity of the dataset, but $K=5$ serves as a reasonable starting point. We observed that using $K=4$ or $K=6$ produced the same imputation results, as shown in Supplementary Table 1.
\begin{figure}[h!]
    \centering
    \includegraphics[width=\textwidth]{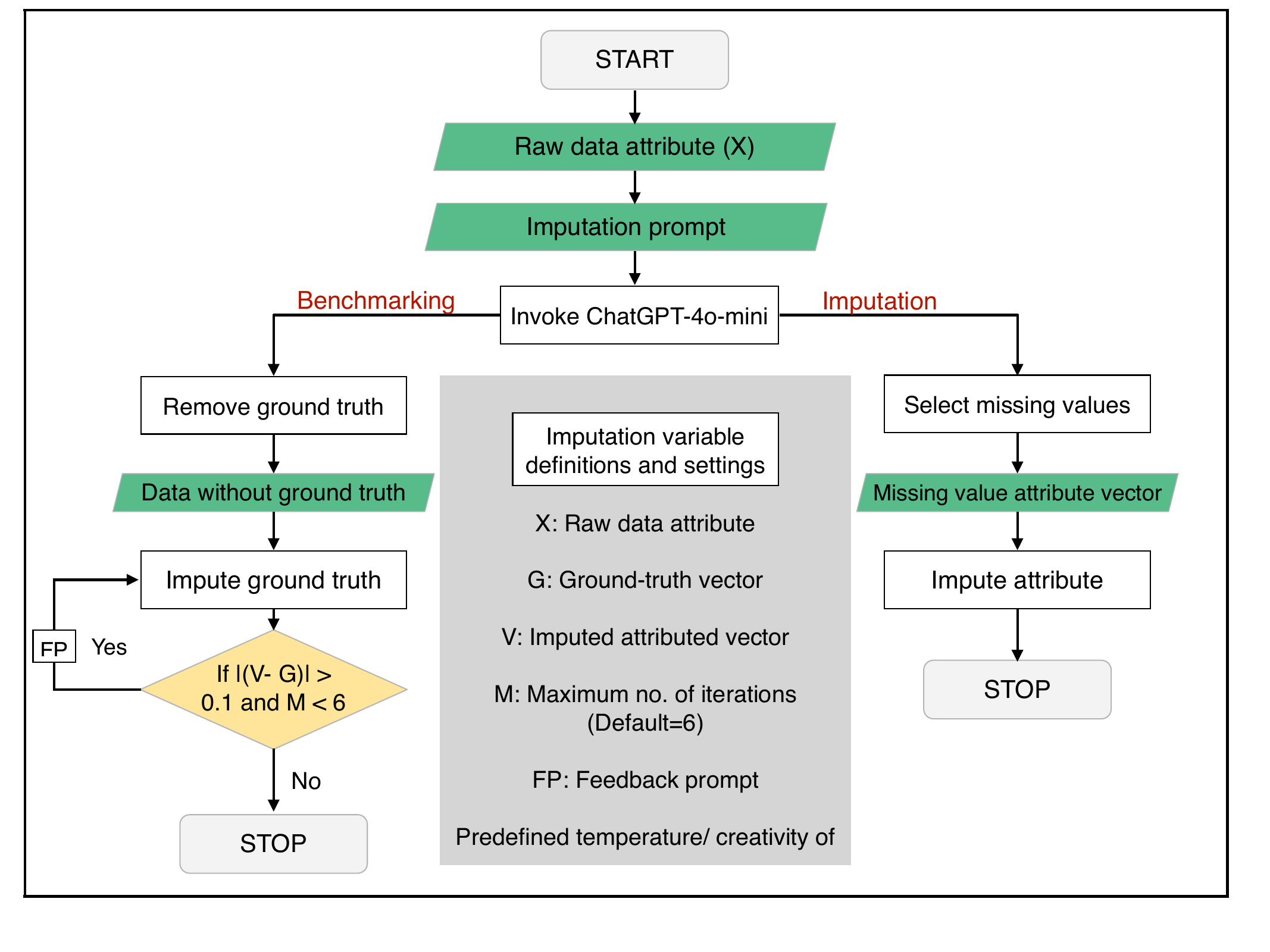}  
    \caption{The workflow of the prompt-based LLM imputation methodology developed in this study.}
    \label{fig:LLMImputation}  
\end{figure}

Besides KNN imputation, we employed prompt-based LLM imputations to handle better sparsely populated or nearly empty attributes, such as the gas flow rates, which did not provide a well-defined data distribution for a distance-based algorithm like KNN. A more ``generally-informed" model, such as an LLM, can leverage syntactical relationships between words to generate data that closely approximate values that might have been reported in the literature. Figure \ref{fig:LLMImputation} shows the prompt-based LLM imputation methodology developed in this study. The prompts used for this task are named based on their general purpose as summarized in Table \ref{tab:promptnames}. Supplementary Note 1.1 details the methodology for each prompt. The GIST prompt introduced here is inspired by an approximation of the CLAIM prompt approach introduced in Ref.\cite{hayat2024claim}. GIST differs from CLAIM as it only incorporates the language descriptors as a pre-processing step to assist the imputation and does not create a missingness-aware description of the data, which is central to the CLAIM methodology. Identifying missingness patterns is not feasible with high-dimensional, low-cardinality data like ours.
The imputed datasets with each method are carried onto the next step of feature engineering. Supplementary Notes 1.3 to 1.5 provide details on imputation methods and their benchmarking. 

LLMs are pre-trained transformer architectures that generate probabilities for the next set of tokens based on an input sequence provided to the model. The next token is predicted as:
\begin{equation}
    t_{\text{next}} = \max_{t} \; P(t \mid \text{context})
\end{equation}
where $t_{\text{next}}$ is the token the model predicts as the next word, $t$ represents each possible next token in the vocabulary, $P(t \mid \text{context})$ is the probability of token $t$ being the next token given the preceding words (the context), and $\max$ denotes selecting the token $t$ with the highest probability. These imputations were performed using the \texttt{ChatGPT-4o-mini} model, which utilizes a lower-dimensional embedding matrix compared to \texttt{ChatGPT-4o} while effectively handling a context window of up to 128,000 tokens. As shown in Figure \ref{fig:LLMImputation}, the imputation strategy using \texttt{ChatGPT} consists of two steps: (a) Benchmarking by reproducing existing ground truth for a selected attribute by first removing its known values and then using the context of other attributes as input to prompt the model to impute removed values; (b) Imputing the missing values for missing attributes based on the learned pattern from step (a). The initial step in imputing data with an LLM like \texttt{ChatGPT} involves selecting an optimal temperature, a parameter that controls the model’s creativity or entropy. OpenAI recommends using a temperature value between 0.2 and 1, depending on the required level of creativity. To determine the most suitable temperature, we calculated the average deviation using \ce{CH4} flow rate imputation. The absolute deviation was calculated for each \ce{CH4} value and then averaged over all available values for each temperature in steps of 0.2. We observed that the lowest average deviation was achieved for the temperature of 0.8, which falls within the OpenAI's recommendation range.
All imputations were therefore performed using a temperature of 0.8. The details of the temperature selection can be found in Supplementary Note 1.2 and Supplementary Figure 2. We introduce a feedback prompt (FP) to step (a), which is triggered when the imputed values from \texttt{ChatGPT} deviate from the respective ground-truth value by a mean absolute error (MAE) greater than 0.1. The number of iterations for this feedback loop is treated as a hyperparameter, with a maximum value of six iterations. The final imputed value is determined by averaging the results over these six iterations according to:
\begin{equation}
    \text{imputed value} = f(\text{Prompt}, \text{LLM model}, \text{context vector}, \text{attribute})
\end{equation}
where LLM model = \texttt{ChatGPT-4o-mini}, context vector = [attributes] - [target], attribute = current attribute being imputed. We also demonstrate a benchmarking experiment where ChatGPT imitates the KNN imputation approach, shown in Supplementary Figure 3. Additional details about the nature of the imputed data points are shown in Supplementary Figures 4 and 5.

\begin{table}[h!]
\centering
\begin{tabularx}{\textwidth}{|X|X|}
\hline
\textbf{Abbreviation} & \textbf{Purpose} \\ \hline
GUIDE: General Instruction for Data Estimation & Imputation based solely on the context of the current row and general instructions \\ \hline
MAP: Multimodal Augmented Prompt              & Extracting instructions from \texttt{ChatGPT-4o} using histogram images of raw data and incorporating them into the prompt for an API call \\ \hline
MAPV: MAP with Variance                       & Modifying MAP by incorporating data variance information \\ \hline
CITE: Contextual Information for Targeted Estimation & Utilizing all available entries of the imputed attribute as context for estimating the missing value \\ \hline
CITEV: CITE with Variance                     & Modifying CITE by incorporating data variance information \\ \hline
GIST: Guide on Integrated Semantic Transformation & A hybrid method that replaces missing values with \texttt{ChatGPT}-generated pre-descriptors before applying the GUIDE prompt \\ \hline
\end{tabularx}
\caption{Abbreviations assigned to various imputation prompts along with their descriptions.}\label{tab:promptnames}
\end{table}

\subsection{Feature Engineering}
\begin{table}[h!]
\centering

\begin{tabularx}{\textwidth}{|X|X|}
\hline
\textbf{\textcolor{black}{Technique}} & \textbf{\textcolor{black}{Summary}} \\ \hline
{One-hot encoding (substrate)} & Leads to too many features \\ \hline 
{Label encoding (substrate)} & Not recommended for too many categories in a classification problem for scarce data \\ \hline 
{Standardization using LLM \& Label encoding (substrate)} & Generates more physically meaningful features with fewer categories, allowing for label encoding \\ \hline 
{Text embedding using LLM (substrate)} & Text embedding model (\texttt{text-embedding-3-small} by OpenAI) generates uniform vectors of specified dimensions for different substrates \\ \hline 
{Label encoding (graphene layers)} & Label encoding for binary and ternary classification \\ \hline 
\end{tabularx}
\caption{Summary of the encoding techniques used to describe the substrate feature in the dataset.} \label{tab:encoding}
\end{table}
We conduct feature engineering to process the substrate attribute, the target variable, and the number of layers. The latter is handled using label encoding, classifying graphene layers into binary categories (monolayer versus multilayer) or ternary categories (monolayer versus bilayer versus multilayer), as the values are discrete. However, for the substrate attribute, as shown in Table \ref{tab:graphene_growth_data}, the data was inconsistently represented across different studies. Labeling each substrate individually would create a high-dimensional feature space, which is undesirable for a sparse dataset. To address this, we utilize OpenAI's text-embedding models, specifically the \texttt{text-embedding-3-small} model, which converts input text into numerical vector representations. This embedding method generates a 128-dimensional vector, which is then reduced to 4- or 8-dimensional vectors using OpenAI's API method, similar to principal component analysis. The text embedding is performed as follows \cite{rong2014word2vec}:
\begin{equation}
    \mathbf{V} = \mathbf{O} \mathbf{E}
\end{equation}

where $\mathbf{O} \in \mathbb{R}^{n \times v}$ is the input vector for $n$ tokens, 
$E \in \mathbb{R}^{v \times d}$ is the embedding matrix and $\mathbf{V} \in \mathbb{R}^{n \times d}$ is the output embedding vector containing the embeddings for all tokens. Dimensions $v$, $n$, and $d$ correspond to the vocabulary size, number of tokens, and size of each token embedding vector, respectively. The implementation of the embedding methodology is elaborated in Supplementary Note 2 and Supplementary Figures 6 and 7. An alternative approach involves using \texttt{ChatGPT} to identify standard descriptors for substrate attributes based solely on the text information provided. This method can function independently but can also be combined with the embedding strategy to enhance feature creation. 
Table \ref{tab:encoding} summarizes the adopted feature engineering techniques, with further details in Supplementary Note 2. Supplementary Table 2 presents examples of standard descriptors generated using \texttt{ChatGPT} and converted into embedding vectors via \texttt{text-embedding-3-small}. The engineered features in the dataset are then discretized using the methods described in the following section. 

\subsection{Discretizion of Feature Space}
\begin{table}[h!]
\centering

\begin{tabular}{|l|p{10cm}|}
\hline
\textbf{Discretization Method} & \textbf{Formulation} \\ \hline
{Equal-width binning} & 
\[
\text{Bin width} = \frac{\text{max} - \text{min}}{n}
\]
\(\text{max}\) = maximum value, \(\text{min}\) = minimum value, \(n\) = number of bins. \\ \hline

{Equal-frequency binning} & 
\[
\text{Entries per bin} = \frac{N}{n}
\]
\(N\) = total number of entries, \(n\) = number of bins. \\ \hline

{K-means binning} & 
\[
\arg \min_S \sum_{i=1}^k \sum_{x \in S_i} \| x - \mu_i \|^2
\]
\(S_i\) = set of points in cluster \(i\), \(\mu_i\) = centroid of cluster \(i\), \(k\) = number of clusters. \\ \hline

{Decision Tree binning} & 
\[
\text{Gini} = 1 - \sum_{i=1}^c p_i^2
\]
\(c\) = number of classes, \(p_i\) = proportion of samples in class \(i\). \\ \hline
\end{tabular}
\caption{Formulations for various feature discretization techniques used in this study.}
\label{tab:table_discretization}
\end{table}
The continuous nature of the attributes can present a significant challenge for models trained on sparse data. To address this limitation, we employ discretization methods, also known as binning techniques, to categorize the features into discrete labels.
Four different discretization techniques are used, including equal width binning, which uniformly divides a range of continuous values into a defined number of bins; equal frequency binning, which ensures that each bin contains the same number of entries based on the sorted data distribution; K-means binning, which groups similar values into $K$ bins, where each bin represents a cluster corresponding to similarly related values with a randomly chosen centroid, where the number of centroids is $K$; and decision tree binning, which splits the data based on the Gini coefficient. Binning outcomes can be mapped to the original data ranges as described in Supplementary Figure 8. For K-means binning, the methodology is shown in Supplementary Figure 10. 
Table \ref{tab:table_discretization} provides a summary of the formulations for the aforementioned discretization techniques, with further details in
Supplementary Note 3.  

\subsection{Machine Learning Models}
The selection of a machine learning model is primarily guided by its ability to generalize patterns from the underlying data distribution rather than simply memorizing them. Effective learning requires balancing model complexity and capacity constraints to ensure meaningful pattern extraction rather than overfitting. However, many machine learning models are susceptible to memorization, often evidenced by low training error but high validation error. Linear models, due to their limited representational power, struggle with complex and diverse datasets, making them poor both at memorization and learning. Conversely, non-linear models, while more expressive, can be prone to overfitting, particularly when trained on imbalanced or scarce data. For instance, tree-based models are highly susceptible to memorization, as they tend to fit each training data point precisely, achieving near-perfect training accuracy but often failing to generalize effectively to unseen data.

Here, we examine four different machine learning models to assess classification performance: Decision Tree, Random Forest, and XGBoost, which are all tree-based models, and Support Vector Machine (SVM), a kernel-based model. SVM has demonstrated a strong learning capacity for classification tasks by separating classes through margin maximization and creating a decision boundary or hyperplane between positive and negative classes. Supplementary Note 4.2 provides details on the SVM formulation and optimization, while Supplementary Note 4 covers additional machine-learning model details.  
%

The collected data, once imputed, feature-engineered, and discretized, serve as training and test datasets for the classification task. In our dataset, about 
30\% and 19\% of data points correspond to bilayer and multi-layer graphene growth, respectively, while the remaining data ($\approx$ 50\%) corresponds to monolayer graphene. We employ stratification when splitting the train-test data, which is a technique to ensure that the distribution of data across different classes of a subspace is representative of the original population (detailed in Supplementary Note 4.3). In this work, we use stratification primarily to create a representative set while calculating the learning curves using a 5-fold cross-validation.

\subsection{Fine Tuning \texttt{ChatGPT}}
\texttt{ChatGPT-4o-mini} was fine-tuned using OpenAI's job creation tool, which allows the specification of training and validation datasets along with some hyperparameters. The goal of fine-tuning is to retrain a few layers of the LLM model such that its responses could be targeted to a desired output format and possibly to draw correlations from the input prompt patterns to predict the output. Specific details on the dataset preparation and fine-tuning models are given in Supplementary Note 5. We choose three epochs to calculate the average cross-validation score, which is translated into token probability scores using a cross-entropy loss formula as described in Supplementary Note 5.
In summary, the fine-tuning job involves (a) dataset preparation, (b) job creation, and (c) API call to the fine-tuned model. The API call retrieves classification results from our fine-tuned models, followed by accuracy report computation, similar to our approach with the SVM model. These fine-tuned models are compared to the SVM model by the accuracy, precision, recall, and F1 score. 
\section{Results}\label{sec:results}
We first evaluate the data imputation performance of all imputation methods by comparing the mean absolute error (MAE) of the imputed data against the true values, as detailed in section \ref{sec:method}. The ground-truth values for each method are derived from the originally complete dataset before introducing artificial missing values. For each attribute, the values that were originally present in the mined dataset, are considered to be the known ground truth values for that attribute and used as references for both benchmarking and imputation steps. KNN imputation benchmark is performed by the statistical average of 50\% known neighbors for the 50\% artificial missing values. The imputed values for the artificial missing placeholders are compared to the ground-truth values, which were removed in the previous step, to compute the deviation. The choice of 50\% neighbors is done to ensure that KNN has representative knowledge before imputation. For LLM-based methods, such as \texttt{ChatGPT} prompt imputation, there is no prior knowledge of neighbors involved. Instead, missing values are imputed based on the model's pre-trained knowledge as well as the information provided through the prompt to guide the imputation. 


\begin{figure}[h!]
    \centering
    \includegraphics[width=\textwidth]{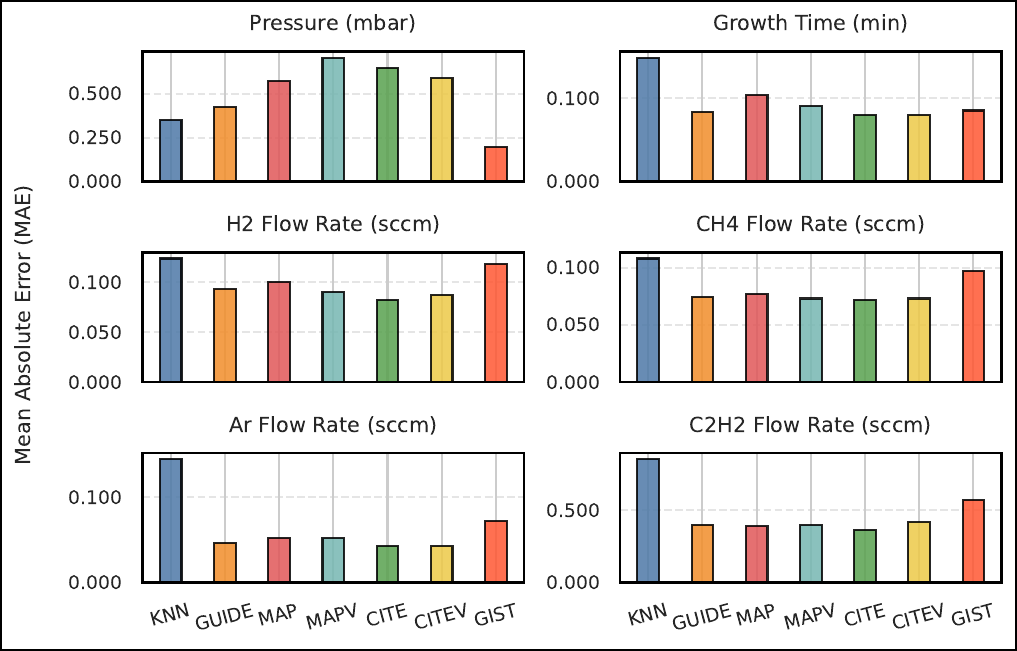}
    \caption{Mean absolute error bars of various imputation techniques used in this study.}
    \label{fig:MAE}
\end{figure}
\begin{figure}[h!]
    \centering
    \includegraphics[width=0.9\textwidth]{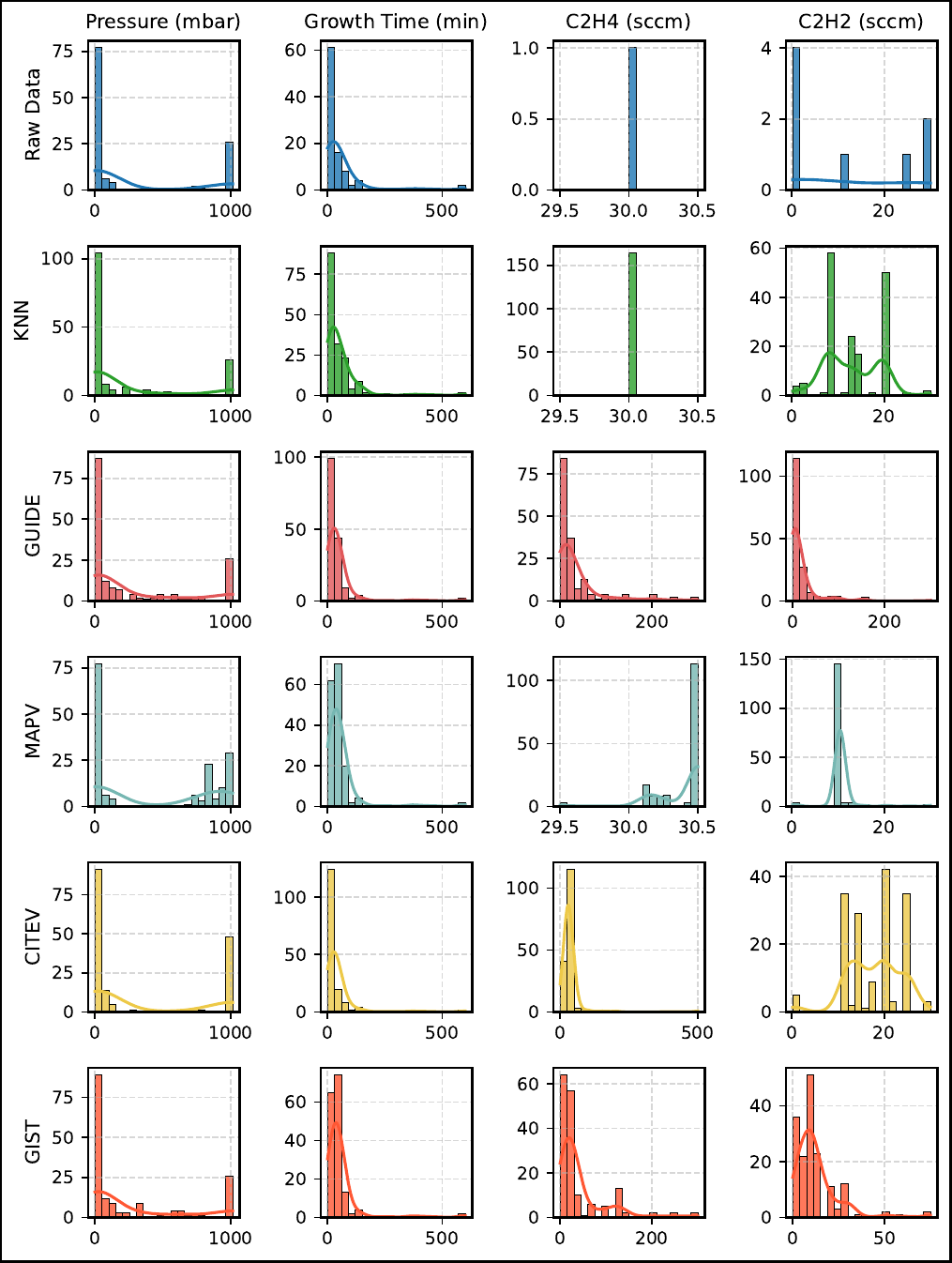}  
    \caption{Comparison of attribute distributions between the existing dataset and the imputed dataset across different imputation methods.}
    \label{fig:distribution}  
\end{figure}

Figure \ref{fig:MAE} compares the MAE for each imputation technique. KNN has the highest MAE for imputing every attribute except pressure.
This is because KNN imputes missing test values based on the similarity of observed data points in the training set. As a result, any test data point with missing values that do not have sufficiently similar neighbors in the training data is likely to be inaccurately imputed by KNN.
This is especially evident in the case of the \ce{C2H4} flow rate, which had only one entry with a value of 30 sccm, forcing KNN to impute 30 sccm for all missing entries (as shown in Figure \ref{fig:distribution}). This observation directly implies that KNN requires a sufficiently large dataset with only a few missing values for effective imputation, whereas sparsely populated attributes are difficult to impute accurately using KNN. All \texttt{ChatGPT} prompts show an improvement in imputation MAE compared to KNN, except for pressure, which is the most populated attribute in the dataset. Among the \texttt{ChatGPT} prompts, GUIDE demonstrates the best overall performance (except for pressure) and the lowest MAE. This is likely due to the generality of the GUIDE prompt, which allows the LLM model to leverage its broad pre-trained knowledge in identifying suitable attribute ranges. The MAPV prompt introduces diversity and data variance instructions during imputation, leading to the highest MAE for pressure compared to all other methods. However, MAPV performs better on sparsely populated attributes, showing lower MAEs than its counterparts, making it a more suitable technique for such cases. The GIST methodology reduces the MAE for pressure compared to GUIDE, representing a slight improvement over GUIDE. Meanwhile, the CITE and CITEV methods yield intermediate MAE values compared to the other prompts.

To evaluate data imputation performance, especially for scarce data, the MAE metric alone is not sufficient and can be misleading when selecting the best imputation technique. An additional metric that quantifies the diversity of the imputed data and its divergence from the original data distribution should be considered. The similarity of the imputed data to the true data distribution is crucial for ensuring realistic imputations. Figure \ref{fig:distribution} compares the distribution of imputed data across different methods against the existing ground-truth data distribution. Due to the scarcity of existing data, there are inherent limitations and biases in representing the underlying distribution of a more comprehensive, unavailable dataset. As shown in Figure \ref{fig:distribution}, this effect is particularly evident for attributes such as \ce{C2H4} and \ce{C2H2} flow rates, where their sparse population causes KNN to closely mimic the original limited-data distribution. In contrast, GPT-based imputation demonstrates some autonomy, allowing it to deviate from the constrained distribution of the original data.
Pressure (mbar) exhibits a bimodal distribution in its raw form, and the imputed data from different methods generally follow this pattern. KNN imputation for pressure closely matches the raw data distribution (as shown in Figure \ref{fig:distribution}). 
The imputation of pressure by GUIDE is slightly more diverse than KNN, with a greater population of imputed values near the first peak. However, MAPV is more inclined toward the second peak, where it concentrates most of the imputed values. This distributional behavior in MAPV is a result of the prompt instructions, which explicitly incorporate both peaks into the structure of the prompt and bias the imputation toward the second peak. On the other hand, CITEV exhibits a distribution similar to KNN, as it closely imitates the raw data distribution by populating most values around the first peak. The contextual information embedded in the prompt allows CITEV to replicate the underlying data structure more precisely, even placing fewer values between the peaks compared to KNN. A more diverse distribution that still aligns with the raw data is produced by the GIST method. GIST imputes pressure and other attributes using pre-defined language descriptors within GUIDE, making it more flexible in generating values in lower-population regions. As shown in Figure \ref{fig:MAE}, GIST exhibits the lowest MAE for pressure imputation. 

An interesting case arises with the \ce{C2H4} flow rate, which has only one entry in the raw data. Since KNN observes only this single value, it imputes the same value for all other rows, resulting in a distribution that is unsuitable for machine-learning tasks. In contrast, the \texttt{ChatGPT} prompts produce more suitable distributions for \ce{C2H4}. This exemplifies how nearly empty attributes can only be effectively handled by LLM-based methods, whereas traditional statistical techniques like KNN fail to generate meaningful variability in such cases. A similar observation can be made for the \ce{C2H2} flow rate, which initially has only eight entries, causing the KNN imputer to deviate from the original distribution. CITEV follows a similar trend but with an even greater deviation. In contrast, the other prompts generate values that follow a more normal distribution, potentially providing a better representation of the general range of this attribute rather than creating artificial peaks, as seen with KNN and CITEV. The imputation of growth time (min) is more consistent across all methods, as they effectively capture the original distribution. This is due to the availability of sufficient data, which ensures that all methods perform well for this attribute.

\begin{figure}[h!]
    \centering
    \includegraphics[width=\textwidth]{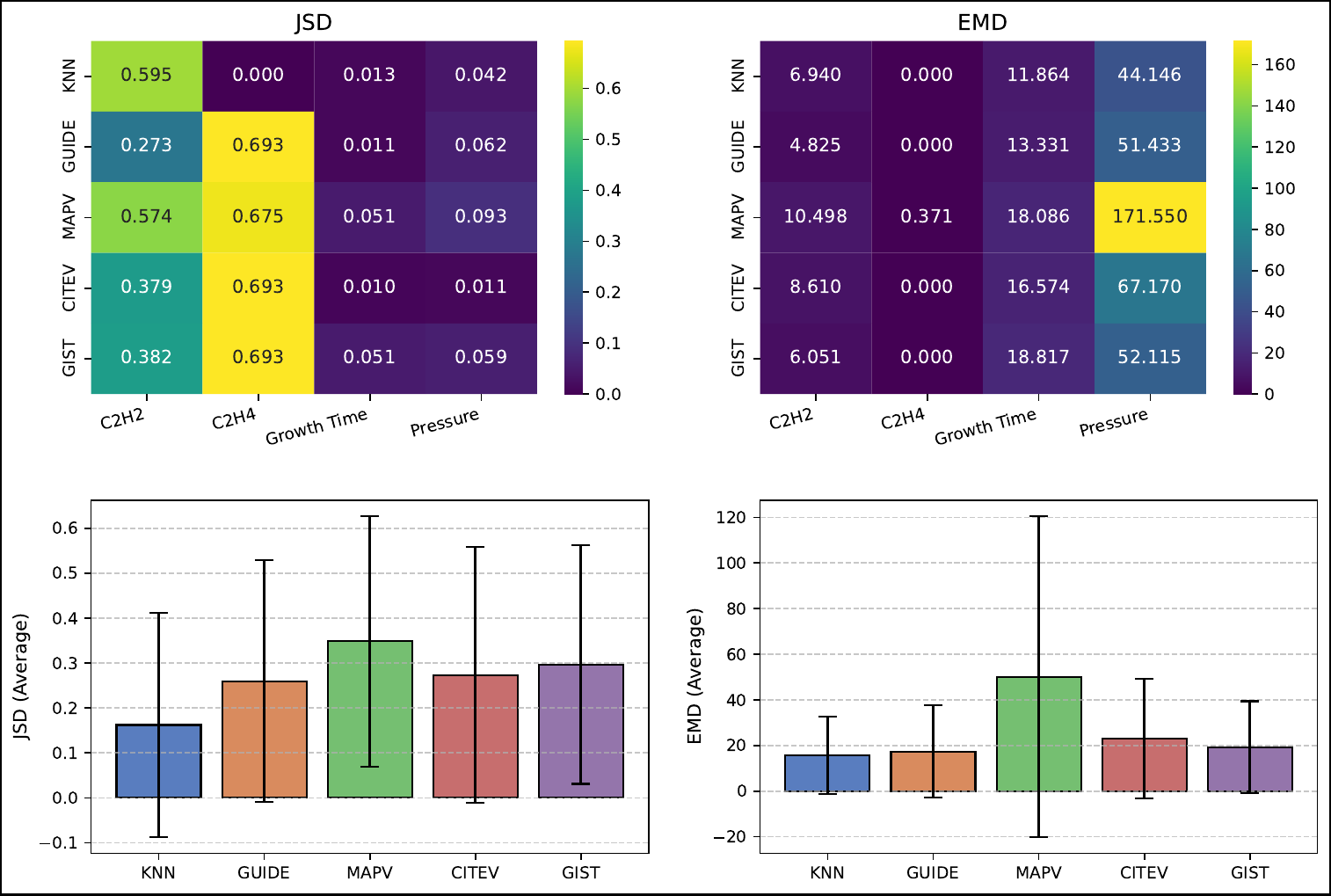}  
    \caption{Comparison of imputation methods based on Jensen-Shannon Divergence (JSD) and Earth Mover's Distance (EMD) across attributes. The top row presents heat maps of different attributes across various imputation methods, while the bottom row depicts the average performance across attributes along with variability.}
    \label{fig:JSD_EMD}  
\end{figure}
The Jensen-Shannon divergence (JSD) and Earth mover's distance (EMD) of the imputed attributes relative to raw data are shown in Figure \ref{fig:JSD_EMD}. JSD measures the similarity between the imputed data and the raw data distribution by quantifying the differences in their shapes relative to a mean distribution. EMD also measures the difference between distributions but additionally accounts for the actual distance between mismatched values. Figure \ref{fig:JSD_EMD} presents the JSD heat map for various attributes across different imputation methods. KNN's low JSD for more populated attributes, such as growth time and pressure, indicates its ability to closely replicate the raw data distribution. Interestingly, KNN exhibits a JSD of zero for the \ce{C2H4} flow rate, as it simply repeats the single available data point across all entries, perfectly replicating the original raw data. In contrast, all \texttt{ChatGPT} prompts show a high JSD for the \ce{C2H4} flow rate, as their imputed distributions deviate from the single-point raw data, regardless of their specific patterns. 

Figure \ref{fig:JSD_EMD} also presents the JSD and EMD values, averaged across all attributes for different imputation methods. Error bars represent the variability of each method across attributes. The MAPV method exhibits the highest JSD from the raw data, highlighting the diverse nature of the imputed data generated by this prompt. In contrast, KNN, GUIDE, and CITEV show lower JSD values, indicating their stronger adherence to the underlying data distribution. The EMD heat map reveals that MAPV also has higher Earth Mover’s Distance (EMD) values, particularly for the pressure attribute, reflecting its bias toward the second peak due to the inclusion of variance information in the prompt. Meanwhile, KNN, GUIDE, and GIST display lower EMD values, suggesting better alignment with the original data distribution. However, the zero EMD values for \ce{C2H4} stem from the lack of data points to measure the distribution difference. Later in this section, we will evaluate the impact of different imputation methods and the resulting imputed data distributions on classification task performance. 
\begin{figure}[h!]
    \centering
    \includegraphics[width=\textwidth]{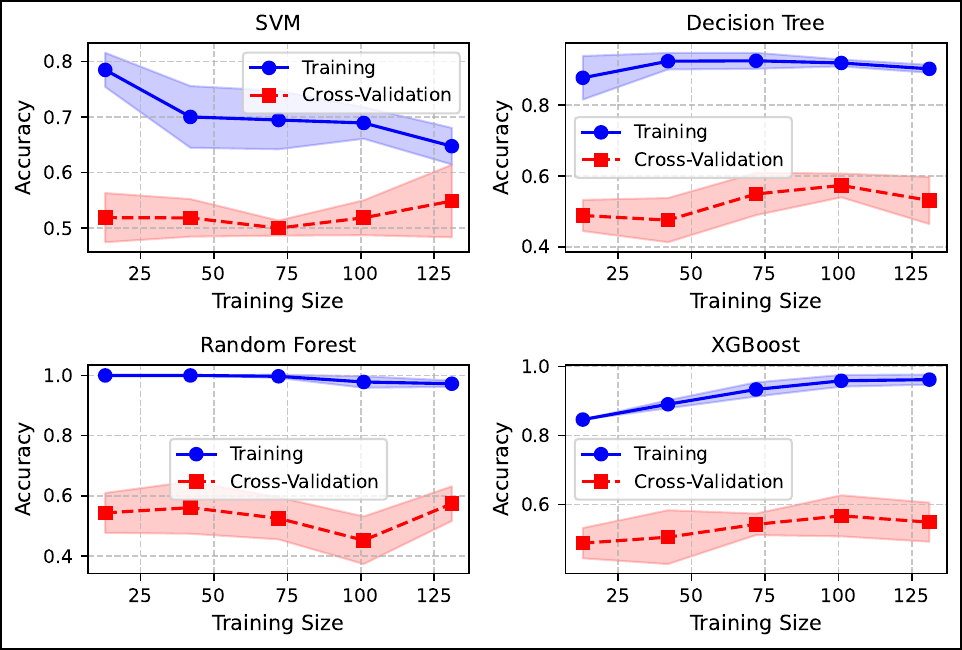}  
    \caption{Learning performance, including classification accuracy on training data and five-fold cross-validation, shown for four different machine learning models.}
    \label{fig:learning}  
\end{figure}

To select the best classification model, we compare the performance of four different models -- Decision Tree, Random Forest, XGBoost, and SVM -- for binary classification on our scarce dataset using the GUIDE imputed dataset with a four-dimensional substrate embedding featurization in the continuous variable space (GUIDE-SE(4)). As shown in Figure \ref{fig:learning}, tree-based models, including Random Forest, Decision Tree, and XGBoost, achieve perfect training scores but exhibit poor cross-validation performance. In contrast, the SVM classifier demonstrates convergence between training and cross-validation scores as the training size increases, highlighting its ability to better learn from scarce data and generalize to unseen instances. The tendency of tree-based models to memorize the training data stems from their higher complexity and expressivity. However, due to the limited data available, they are prone to overfitting, as indicated by the large gap between their training and cross-validation curves. Another key observation is that all models are constrained by the limitations of the scarce dataset. While SVM exhibits the least overfitting and achieves the highest validation accuracy of around 60\%, the cross-validation scores of all models generally fall between 50\% and 60\%. This suggests that, regardless of the model architecture, the overall learning process is fundamentally restricted by data scarcity.


\begin{figure}[htp]
    \centering
    \includegraphics[width=0.95\textwidth]{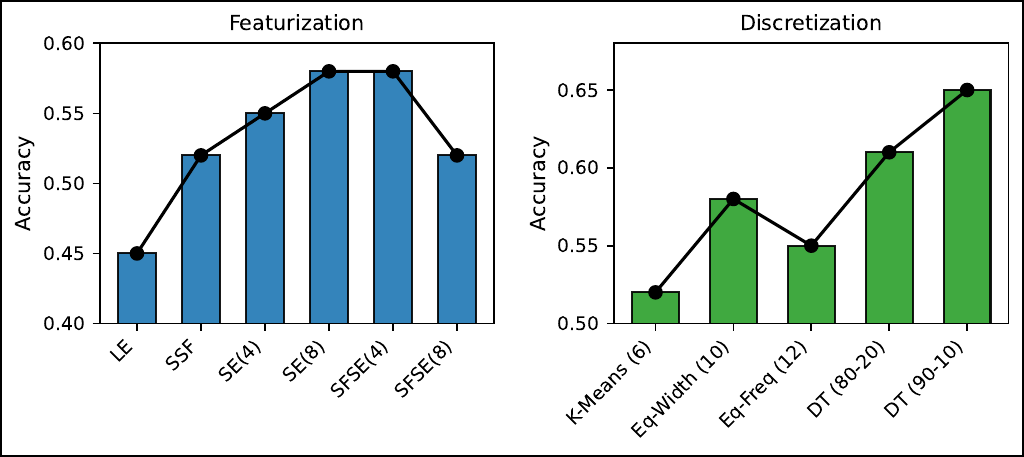}  
    \caption{The binary classification (monolayer versus multi-layer) accuracy of the SVM model on GUIDE imputed data for different substrate featurization and discretization methods.}
    \label{fig:featurization_discretization}  
\end{figure}

To assess the impact of different substrate featurization methods, as detailed in Table \ref{tab:encoding}, on learning performance, we compare the accuracy of an SVM binary classifier (monolayer vs. multilayer graphene) on GUIDE-imputed data across various featurization approaches. Figure \ref{fig:featurization_discretization} illustrates the classification accuracy for each method. The classifier performs similarly to a naive model (accuracy < 0.5) when using label-encoded (LE) features. However, the LLM-standardized substrate features (SSF) without embeddings improve accuracy from 45\% to 52\% compared to LE (examples of SSF are shown in Supplementary Table 2). Additionally, applying the \texttt{text-embedding-3-small} model to the raw substrate attributes yields higher accuracy, particularly when using 4-dimensional (SE(4)) and 8-dimensional (SE(8)) embedding vectors. We further vectorize the key-value pairs of the standardized substrate features using the embedding model to evaluate the accuracy, referred to as Standardized Features on Substrate Embedding (SFSE). The SFSE(4) method achieves the same accuracy as SE(8), but increasing the dimensions further to SFSE(8) results in lower accuracy. This decline is likely due to the trade-off between feature space dimensionality and data population. SE(8) effectively captures meaningful and contextual representations by leveraging the embedding model, extending into higher dimensions. However, for SFSE(8), the variability in input representation is lower (i.e., due to standardization), leading to sparse contextual embeddings that primarily reflect the predefined contexts in the key-value pairs. Classification performance metrics for different featurization methods are detailed in Supplementary Table 3. 

We also investigated the impact of discretizing the feature space on classification performance, as discretization can be beneficial in data-scarce scenarios. Figure \ref{fig:featurization_discretization} compares the accuracy of the SVM binary classifier for GUIDE-imputed data with the SE(4) substrate featurization across different discretization approaches, as detailed in Table \ref{tab:table_discretization}.
K-means binning with six clusters (K-Means(6)) slightly reduces accuracy compared to SE(4) in the continuous space. This decline is likely due to suboptimal bin splits, as K-means minimizes intracluster variance, leading to non-uniform bin divisions. In contrast, equal-width binning with ten bins (Eq-Width(10)) yields higher accuracy than both K-Means(6) and continuous SE(4). Eq-Width(10) divides attributes into ten equally spaced bins, ensuring uniform discretization by definition, which results in improved accuracy over K-means-based binning.
Conversely, equal-frequency binning with twelve bins (Eq-Freq(12)) leads to lower accuracy, likely due to the same issue of suboptimal or nonuniform bin splits. The effect of different bin sizes is discussed in Supplementary Figure 9. Additionally, we examine a supervised discretization approach using decision trees (DT), which optimizes bin splits by minimizing the Gini impurity, ensuring bins are formed based on attribute-target correlations. This method is particularly effective, as it clusters bins in a more informed manner than unsupervised binning methods. The accuracy of an 80-20 train-test split with DT (DT(80-20)) is significantly higher than that of unsupervised binning methods and is further improved with a 90-10 split (DT(90-10)). The classification performance metrics for different discretization methods are reported in Supplementary Table 4.
\begin{figure}[h!]
    \centering
    \includegraphics[width=\textwidth]{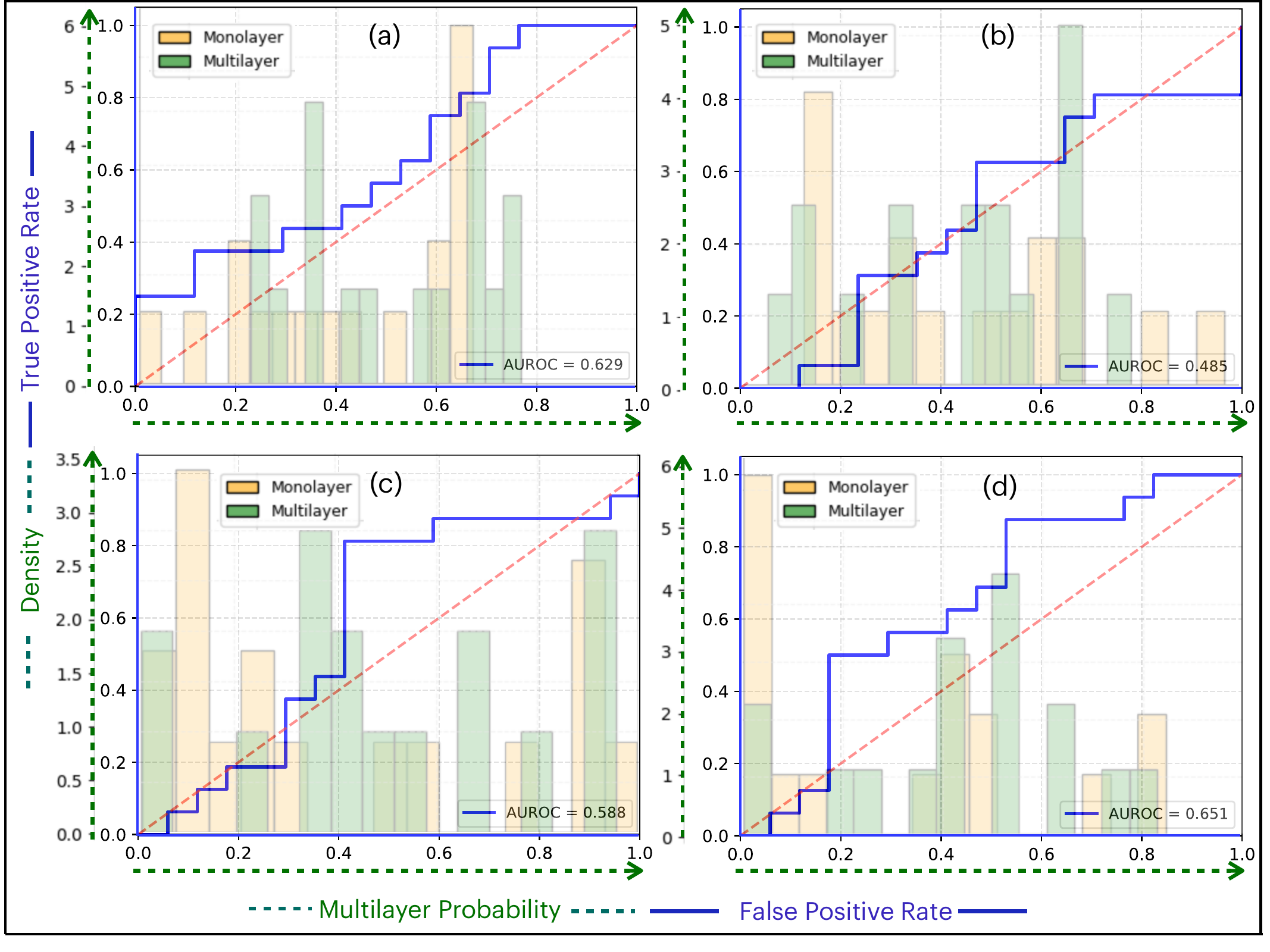}  
    \caption{Impact of different data imputation methods on SVM binary classification performance, for SE(4) substrate featurization in the continuous variable space using (a) GUIDE, (b) KNN, (c) MAPV, and (d) CITEV imputation methods.} 
    \label{fig:imputation_on_learning}  
\end{figure}

We further investigate the impact of different data imputation methods on SVM binary classification performance using SE(4) substrate featurization in the continuous variable space. Figure \ref{fig:imputation_on_learning} compares the Area Under the Receiver Operating Characteristic Curve (AUROC) values for the SVM binary classifier trained on different imputed datasets, as well as the distribution density of the test samples. The AUROC value indicates a classifier's ability to distinguish between positive and negative classes, making it a more informative metric than accuracy alone. An AUROC (or AUC) value of 1 represents a perfect classifier, corresponding to a completely filled square in the ROC space.
As shown in Figure \ref{fig:imputation_on_learning}, GUIDE achieves an AUC score of 63\% with minimal overfitting, as evidenced by the small gap between cross-validation and training scores shown in Supplementary Figure 11. In contrast, KNN exhibits a significant drop in AUC score to 49\% and shows notable overfitting, indicated by the larger gap between training and cross-validation scores (Supplementary Figure 11). This suggests that KNN-imputed data is not well-suited for classification tasks on scarce datasets. MAPV yields an intermediate AUC score of 59\%, positioned between GUIDE and KNN. Notably, GUIDE, which does not incorporate data-specific instructions in its input prompt, outperforms MAPV, which includes such instructions. This implies that \texttt{ChatGPT}'s pre-trained knowledge may be sufficient to capture the range of imputed attributes, making GUIDE more effective than MAPV in this context. Results for MAP and CITE, presented in Supplementary Figure 12, indicate that both methods suffer from overfitting. While CITEV achieves the highest AUC, it does so at the expense of overfitting the model. This is likely because the prompt includes the full attribute context during imputation, leading to replication of the underlying data. However, in CITEV, the introduction of variability instructions improves AUC compared to KNN-imputed data. These findings suggest that enhancing diversity in the originally scarce and constrained dataset is more beneficial for improving classification accuracy than simply using statistical tools to replicate the underlying distribution, which may carry biases associated with data scarcity.

\begin{figure}[h!]
    \centering
    \includegraphics[width=\textwidth]{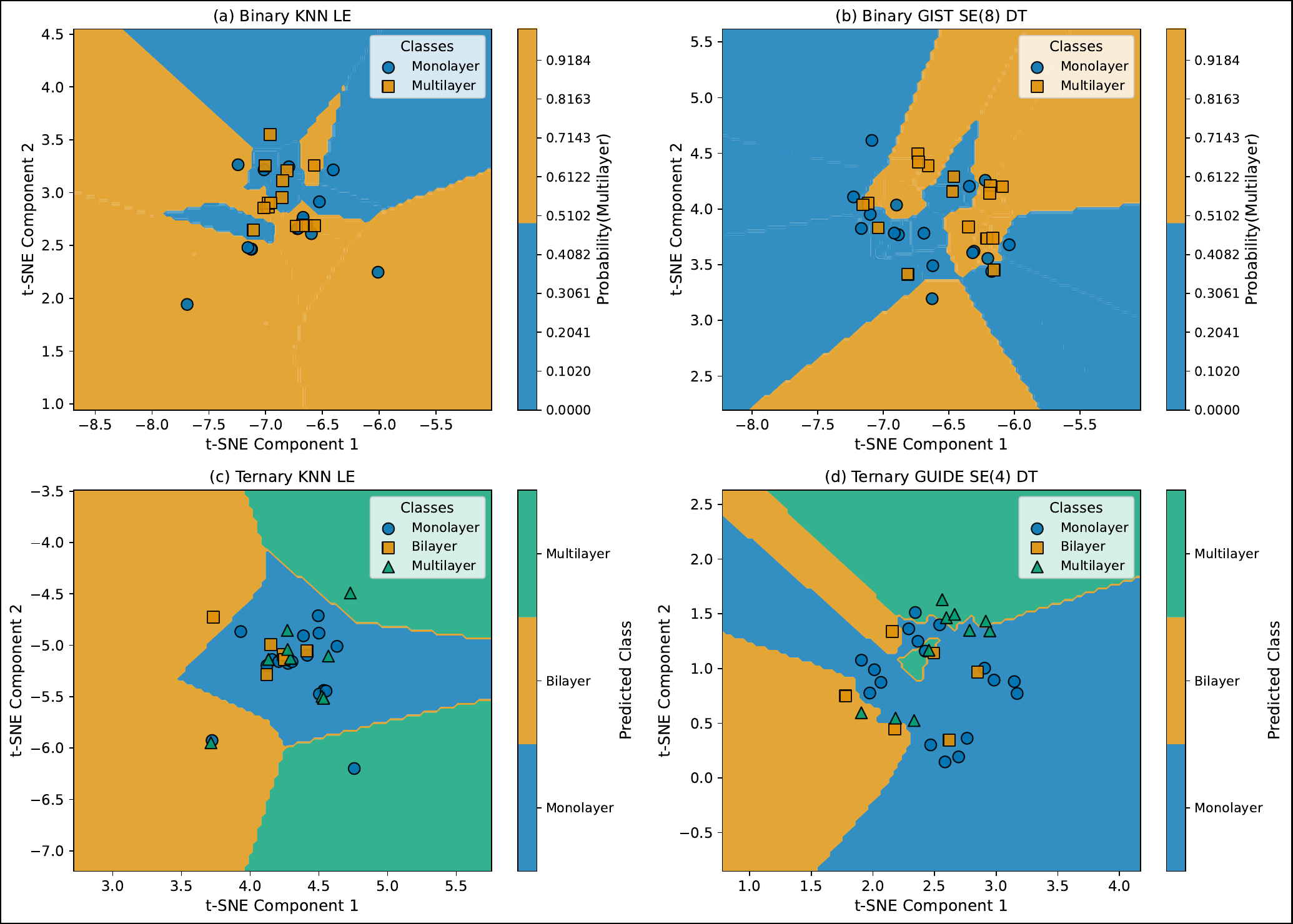}  
    \caption{t-SNE visualization of the SVM classifier for (a) KNN-LabelEncoded-Continuous and (b) GIST-SE(8)-DT in binary classification, as well as (c) KNN-LabelEncoded-Continuous and (d) GUIDE-SE(4)-DT in ternary classification.}
    \label{fig:worst_best}  
\end{figure}

Finally, we evaluate the cumulative impact of all the strategies presented in this study for improving classification performance on scarce, inhomogeneous data. To do so, we compare the most basic data processing and feature engineering approach against the most effective strategy identified in this study for the SVM classifier. The least processed approach consists of KNN-imputed data with label-encoded substrate featurization in the continuous feature space (KNN-LabelEncoded-Continuous). In contrast, the most advanced method utilizes GIST-imputed data with text-embedding-based substrate featurization (8-dimensional) and decision tree discretization (GIST-SE(8)-DT). The exception to this is for the ternary case where the GUIDE-imputed data with four-dimensional substrate embedding and decision tree discretization shows the highest accuracy (GUIDE-SE(4)-DT). Figure \ref{fig:worst_best} illustrates t-distributed stochastic neighbor embedding (t-SNE) maps for the binary SVM classifier, comparing the worst-performing (KNN-LabelEncoded-Continuous) and best-performing (GIST-SE(8)-DT) models. For the ternary case, it shows the comparison between the worst-performing (KNN-LabelEncoded-Continuous) and the best-performing (GUIDE-SE(4)-DT). These results show a significant accuracy improvement from 39\% to 65\% for binary classification (Figure \ref{fig:worst_best}(a)-(b)) and from 52\% to 72\% for ternary classification (Figure \ref{fig:worst_best}(c)-(d)).

The t-SNE representation of Figure \ref{fig:worst_best} also illustrates the probability of different data points being correctly classified. The best-performing models establish new decision boundaries that more effectively separate classes. 
The strategies presented in this study, along with the step-by-step approach outlined here, are readily applicable for enhancing machine learning performance on scarce datasets in a broader context, where similar challenges often lead to low accuracy.

\begin{figure}[h!]
    \centering
    \includegraphics[width=0.7\textwidth]{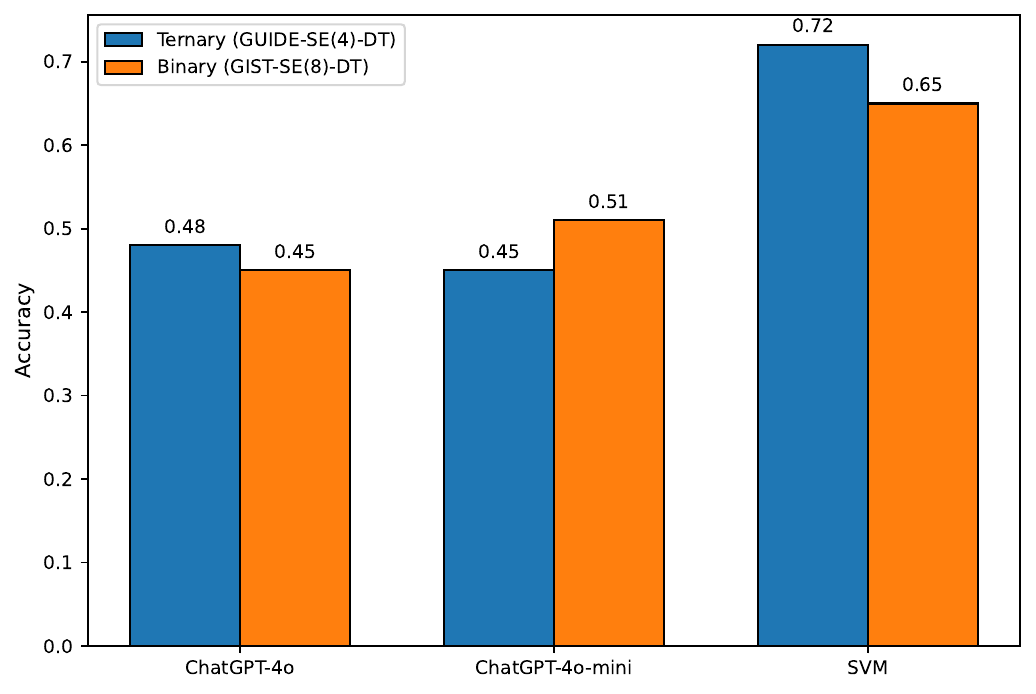}  
    \caption{Accuracy comparison of the best binary and ternary classifiers between two fine-tuned GPT models and the SVM model.}
    \label{fig:fine_tune_compare}  
\end{figure}
Additionally, we fine-tune two \texttt{ChatGPT-4o} models to directly classify the binary and ternary classes and compare their performance with the SVM classifier. The same training datasets were used to fine-tune the models. The technical details are presented in Supplementary Note 5. The predictions of both fine-tuned \texttt{ChatGPT-4o} and \texttt{ChatGPT-4o-mini} are shown in Figure \ref{fig:fine_tune_compare} along with the SVM classifier. The LLM-based classification is significantly outperformed by the more standard SVM classifier for both binary and ternary cases. The scores reflect the performance of both approaches under scarce data conditions, showing that traditional ML models combined with LLM-based data engineering techniques outperform fine-tuned LLM models alone. Detailed accuracy reports are provided in Supplementary Table 5, and the cross-validation curve for fine-tuning process is shown in Supplementary Figure 13. The potential of an LLM fine-tuned on larger datasets remains speculative and is beyond the scope of this study. Our general guidelines can leverage LLM to improve conventional data engineering strategies to optimize learning for numerical models.

\section{Conclusion}
The integration of LLMs can significantly enhance machine learning models trained on scarce and heterogeneous datasets. Statistical techniques like KNN are best suited for imputing data in well-populated datasets, allowing the algorithm to learn from the underlying data distribution. However, LLMs offer a more versatile approach, as they can impute data regardless of missing data density. This study shows that prompt-based LLM techniques yield lower errors in generating data distributions than KNN while effectively reproducing ground truth, consistent with previous findings \cite{he2024llm, wang2025llm, hayat2024claim, ji2024predicting}. We showcased that the performance of the SVM classifier improves in both accuracy and AUROC for binary and ternary classification when LLM-based featurization and discretization techniques are applied. Using the proposed methodology, a naive binary classifier trained on scarce data with an initial accuracy of 46\% can be enhanced to 65\% accuracy through readily accessible LLM-based prompt techniques. These results underscore the importance of data engineering in significantly improving machine learning performance, particularly in data-scarce scenarios. The developed models serve as preliminary estimators for graphene synthesis classification, with potential for future refinement using techniques like $\Delta$ machine learning.


\label{sec:conclusion}

\section*{Acknowledgment} 
This research was supported by the National Science Foundation (NSF) under Award Number DMR-2119308. We utilized resources at the Electronic Visualization Laboratory (EVL) at the University of Illinois Chicago, made available through NSF Award CNS-1828265. The authors extend their gratitude to Professor Michael Trenary for his valuable discussions and guidance in collecting graphene growth data.

\section*{Author Contribution Statement} 
D.B. and S.K. conceived the research. S.K. supervised the study and provided financial support. D.B. developed the methodology, performed the calculations, and performed the analysis. Both authors contributed to the discussion of the results and the editing of the manuscript.

\section*{Competing Interests}
The authors declare that they have no competing financial or non-financial interests.

\section*{Data Availability Statement}

Data and methodology used to produce the results for this study are presented in the Supplementary file and in the GitHub repository \url{https://github.com/kadkhodaei-research-group/GScarceLLM.git}. 
\bibliography{bib.bib}

\cleardoublepage

\renewcommand{\thesection}{Supplementary Note \arabic{section}}
\renewcommand{\thefigure}{Supplementary Figure \arabic{figure}}
\renewcommand{\figurename}{}
\renewcommand{\thetable}{Supplementary Table \arabic{table}}
\renewcommand{\tablename}{}  
\renewcommand{\theequation}{S\arabic{equation}}

\setcounter{section}{0}
\setcounter{figure}{0}
\setcounter{table}{0}
\setcounter{equation}{0}







\clearpage  
\onecolumn  

\begin{center}
    {\LARGE {Supplementary Material}} \\[1em]
    {\large for} \\[1em]
    {\LARGE {Leveraging Large Language Models to Address Data Scarcity in Machine Learning: Applications in Graphene Synthesis}} \\[1em]
    {\large Devi Dutta Biswajeet$^1$ and Sara Kadkhodaei$^1$} \\[1em]
    {\small $^1$Department of Civil, Materials, and Environmental Engineering, University of Illinois Chicago, Chicago, IL, United States of America}
\end{center}

\section{Data Imputation}
The raw dataset contains data entries for attributes corresponding to the graphene growth conditions and graphene number of layers, i.e., the target variable. The raw attributes are Pressure (mbar), Temperature (°C), Growth Time (min), Substrate,	Number of Graphene Layers, and gas Flow Rates (sccm). These attributes and their corresponding entries are mined from the experimental chemical vapor deposition (CVD) graphene growth literature \cite{saeed2020chemical, hong2023recent, son2017low, liu2011synthesis, zhang2013review, liu2017controlled, bhuyan2016synthesis, chen2015large, huang2021substrate}. The gas flow rates are initially extracted as a dictionary- \{gas: flow rate\}, later separated into different columns. Out of many gases reported in the literature, we select the top 5 most common gases -- H2, CH4, C2H4, Ar, and C2H2. This is to keep the number of attributes to a minimum, which is especially important to address the curse of dimensionality for scarce datasets. 
The Supplementary File ``DataProcessing.ipynb'' shows the count of each attribute before the imputation process and can be used to reproduce the results for the imputation step. The collected data are stored in the ``graphene\_growth\_conditions\_layers.csv'' file.

\begin{figure}[h!]
    \centering
    \includegraphics[width=0.8\textwidth]{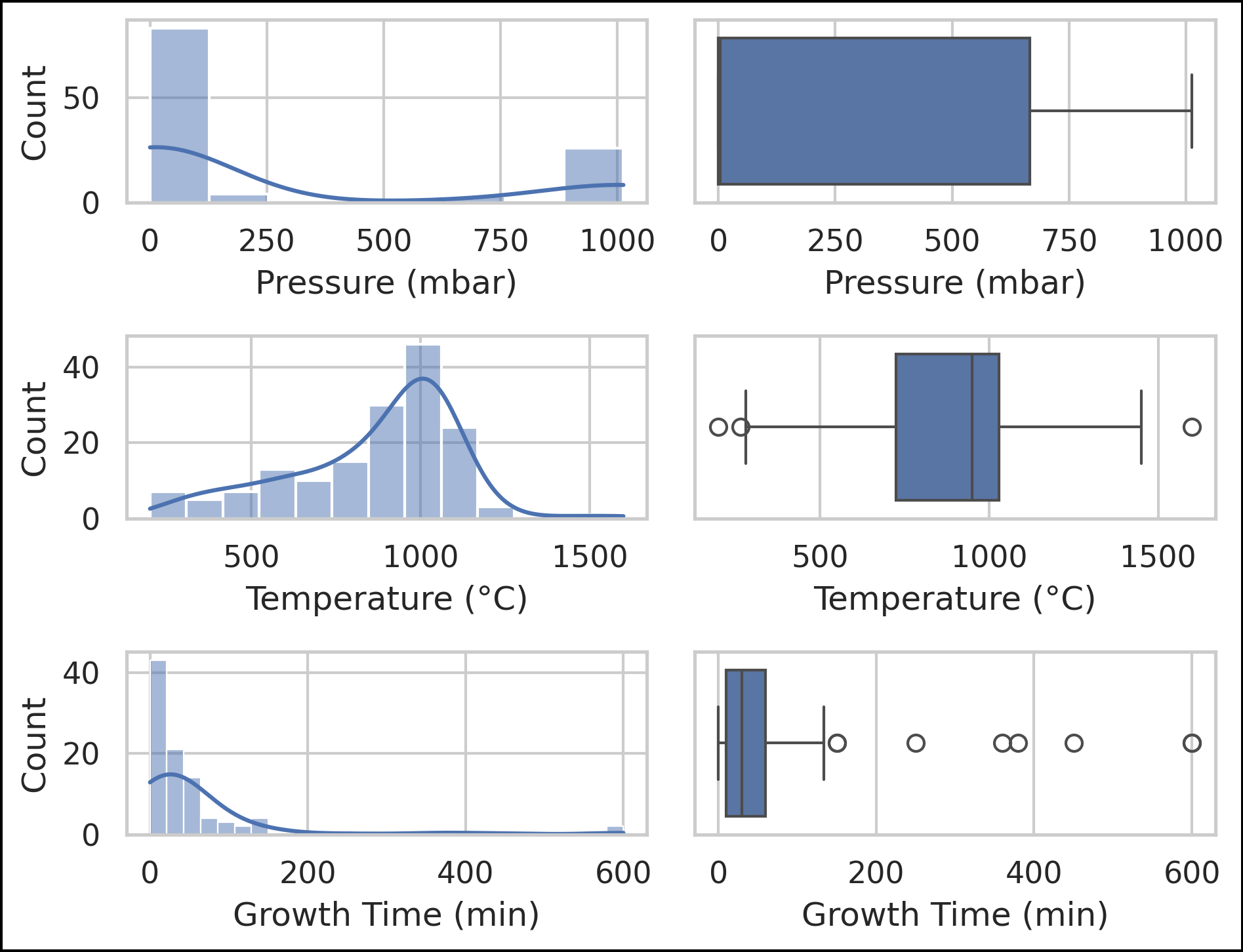}  
    \caption{Histogram and box plots showing the raw data distribution for Pressure (mbar), Temperature ($^{\circ}$C), and Growth Time (min).}
    \label{fig:histograms}
\end{figure}

The histogram and box plot distributions for Pressure (mbar), Temperature (°C), and  Growth Time (min) are shown in \ref*{fig:histograms}.
After the preprocessing step of extracting all the raw features, the imputation step is implemented with two techniques: \( (a) \) K-Nearest Neighbors (KNN) and \( (b) \) Large Language Model (LLM), specifically \texttt{ChatGPT-4o-mini}.
The KNN imputation is performed by selecting the top 5 neighbors for the missing attribute by calculating the Euclidean distance between the other known attributes and then using the available values of the original attribute from the nearest neighbors to impute with the mean value. The implementation is given in ``DataProcessing.ipynb".\\ 

\subsection{ChatGPT Prompts}
There is no universal strategy to prompt engineering as LLMs tend to be a dynamic model that learns from data regularly. However, certain key elements of a prompt involve the chain-of-thought imposed on the language model to address challenges like hallucination. We use a simple prompt strategy involving the description of the \textbf{System} and \textbf{User} placeholders analogous to global and local instructions. We provide general instructions of what the LLM should act like for the \textbf{System} placeholder, for example, an imputation expert which utilizes the knowledge from literature of CVD processes. The \textbf{User} placeholder gets instructions on what to do based on its capability, for example, imputing the value of an attribute. The instructions depend on the requirement of the structured output, which can be specified at the end. The hallucination part is handled using simple instructions within the \textbf{System} argument. The description of the prompts is given below.
\\

\noindent {\textbf{1. GUIDE}}: \\\\
\texttt{\textcolor{blue}{\textbf{System:}} You are a CVD expert with access to experimental and/or simulated data from the literature, which you can understand to derive correlations between known and unknown quantities for given datasets.\\ Now, I would like you to help me impute the missing value for \{target\_variable\}.\\ Please use your scientific knowledge to impute this value, based only on the current row's context. Note that the values provided are not normalized, and the imputation should be done using your understanding of CVD processes. }\\
\texttt{\textcolor{red}{\textbf{User:}} For the CVD process '\{context\_vars['cvd\_method']\}', with the following parameters: \\ Pressure: \{context\_vars['pressure']\} mbar, Temperature: \{context\_vars['temperature']\} °C, Growth Time: \{context\_vars['growth\_time']\} min, Substrate: \{context\_vars['substrate']\}, Gas Flow Rates: H2: \{context\_vars['h2']\}, C2H4: \{context\_vars['c2h4']\}, Ar: \{context\_vars['ar']\}, C2H2: \{context\_vars['c2h2']\}. Can you determine the missing value of \{target\_variable\}? Please provide only the missing value as a single number without any additional text. \\\\}

\noindent {\textbf{2. MAP}}: \\\\
\texttt{\textcolor{blue}{\textbf{System:}} You are a CVD expert with access to experimental and/or simulated data from literature, which you can understand to derive correlations between known and unknown quantities for given datasets. Now, I would like you to help me impute the missing value for \{target\_variable\}. Please use your scientific knowledge to impute this value based only on the current row's context. \\
Additionally, ensure that the imputation for \{target\_variable\} follows the original data distribution, as observed in the dataset for each parameter. Here are the key characteristics of the distributions:
\begin{itemize}
    \item \textbf{C2H4}: Almost all values are exactly 30, indicating a single known value without much variability.
    \item \textbf{Pressure (mbar)}: Skewed right, with the majority of the data close to 0 and a second small peak around 1000 mbar.
    \item \textbf{Growth Time (min)}: Skewed right with most values concentrated below 100 min, though there are a few outliers up to 600 min.
    \item \textbf{H2}: Highly right-skewed, with most values near zero and a few extending to higher values.
    \item \textbf{CH4}: Right-skewed, with most values concentrated near 0.
    \item \textbf{Ar}: Very right-skewed, with most values concentrated near zero and very few extending up to 10,000.
    \item \textbf{C2H2}: Multimodal, with peaks near 0, 10, and 30.
\end{itemize}
Note that the values provided are not normalized, and the imputation should be done using your understanding of CVD processes and the data distribution patterns observed above. \\
\textcolor{red}{\textbf{User:}} For the CVD process '\{context\_vars['cvd\_method']\}', with the following parameters:
\begin{itemize}
    \item Pressure: \{context\_vars['pressure']\} mbar
    \item Temperature: \{context\_vars['temperature']\} °C
    \item Growth Time: \{context\_vars['growth\_time']\} min
    \item Substrate: \{context\_vars['substrate']\}
    \item Gas Flow Rates: H2: \{context\_vars['h2']\}, C2H4: \{context\_vars['c2h4']\}, Ar: \{context\_vars['ar']\}, C2H2: \{context\_vars['c2h2']\}
\end{itemize}
Can you determine the missing value of \{target\_variable\}? Please provide only the missing value as a single number without any additional text.\\}

\noindent {\textbf{3. MAPV}}: \\\\
\texttt{\textcolor{blue}{\textbf{System:}} You are a CVD expert with access to experimental and/or simulated data from literature, which you can understand to derive correlations between known and unknown quantities for given datasets. Now, I would like you to help me impute the missing value for \{target\_variable\}. Please use your scientific knowledge to impute this value, based only on the current row's context, and ensure that the imputation follows the original data distribution while introducing controlled variability.} \\

\texttt{The original data distributions for each attribute are as follows:
\begin{itemize}
    \item \textbf{C2H4}: Tightly concentrated around 30 with minimal variability.
    \item \textbf{Pressure (mbar)}: Right-skewed, with most values at low pressures and a secondary peak around 1000 mbar.
    \item \textbf{Growth Time (min)}: Right-skewed, with most values below 100 minutes and some outliers extending to 600 minutes.
    \item \textbf{H2}: Right-skewed, with values concentrated near 0, but with a long tail up to 1400.
    \item \textbf{CH4}: Right-skewed, with most values near 0 and a few outliers extending up to 500.
    \item \textbf{Ar}: Highly right-skewed, with most values near 0 and some outliers extending up to 10,000.
    \item \textbf{C2H2}: Multimodal, with distinct peaks around 0, 10, and 30.
\end{itemize}
}
\texttt{In order to introduce realistic variability and avoid repetition of imputed values, a small amount of random noise should be applied uniformly to all attributes during imputation. This noise should be small enough to maintain the general data distribution but provide natural variations, ensuring that no extreme outliers are introduced. The noise should be probabilistically scaled based on each attribute's distribution, with skewed distributions respecting their inherent shape while still introducing variability.}

\noindent\texttt{\textcolor{red}{\textbf{User:}} For the CVD process '\{context\_vars['cvd\_method']\}', with the following parameters:
\begin{itemize}
    \item Pressure: \{context\_vars['pressure']\} mbar
    \item Temperature: \{context\_vars['temperature']\} °C
    \item Growth Time: \{context\_vars['growth\_time']\} min
    \item Substrate: \{context\_vars['substrate']\}
    \item Gas Flow Rates: H2: \{context\_vars['h2']\}, C2H4: \{context\_vars['c2h4']\}, Ar: \{context\_vars['ar']\}, C2H2: \{context\_vars['c2h2']\}
\end{itemize}
Can you determine the missing value of \{target\_variable\}? Please provide only the missing value as a single number, without any additional text.\\}

\noindent{\textbf{4. CITE}}: \\\\
\texttt{\textcolor{blue}{\textbf{System:}} You are a CVD expert with access to experimental and/or simulated data from literature, which you can understand to derive correlations between known and unknown quantities for given datasets. \{analysis\_message\} Now, I would like you to help me impute the missing value for \{target\_variable\}. Please use your scientific knowledge to impute this value, based only on the current row's context. Note that the values provided are not normalized, and the imputation should be done using your understanding of CVD processes. }

\noindent\texttt{\textcolor{red}{\textbf{User:}} For the CVD process '\{context\_vars['cvd\_method']\}', with the following parameters:
\begin{itemize}
    \item Pressure: \{context\_vars['pressure']\} mbar
    \item Temperature: \{context\_vars['temperature']\} °C
    \item Growth Time: \{context\_vars['growth\_time']\} min
    \item Substrate: \{context\_vars['substrate']\}
    \item Gas Flow Rates: 
    \begin{itemize}
        \item H2: \{context\_vars['h2']\}
        \item C2H4: \{context\_vars['c2h4']\}
        \item Ar: \{context\_vars['ar']\}
        \item C2H2: \{context\_vars['c2h2']\}
    \end{itemize}
\end{itemize}
Can you determine the missing value of \{target\_variable\}? Please provide only the missing value as a single number without any additional text. \\
Here's the current attribute data, which you need to impute based on this knowledge and your knowledge of the literature: \{data\_preprocessed[target\_variable].to\_string()\} \\} %

\noindent {\textbf{5. CITEV}}: \\\\
\texttt{\textcolor{blue}{\textbf{System:}} You are a CVD expert with access to experimental and/or simulated data from literature, which you can understand to derive correlations between known and unknown quantities for given datasets. \{analysis\_message\} Now, I would like you to help me impute the missing value for \{target\_variable\}. Please use your scientific knowledge to impute this value, based only on the current row's context. Note that the values provided are \textbf{not normalized}, and the imputation should be done using your understanding of CVD processes. }

\noindent\texttt{\textcolor{red}{\textbf{User:}} For the CVD process '\{context\_vars['cvd\_method']\}', with the following parameters:
\begin{itemize}
    \item Pressure: \{context\_vars['pressure']\} mbar
    \item Temperature: \{context\_vars['temperature']\} °C
    \item Growth Time: \{context\_vars['growth\_time']\} min
    \item Substrate: \{context\_vars['substrate']\}
    \item Gas Flow Rates:
    \begin{itemize}
        \item H2: \{context\_vars['h2']\}
        \item C2H4: \{context\_vars['c2h4']\}
        \item Ar: \{context\_vars['ar']\}
        \item C2H2: \{context\_vars['c2h2']\}
    \end{itemize}
\end{itemize}
Can you determine the missing value of \{target\_variable\}? Please provide only the missing value as a single number, without any additional text. \\ Here's the current attribute data, you need to impute based on this knowledge and your knowledge of the literature:
Also make sure to impute values around the modes and also in regions of scarely populated data, to have a diverse data distribution
\{data\_preprocessed[target\_variable].dropna().iloc[:len(data\_preprocessed[target\_variable].dropna()) // 2].to\_string()\\}

\noindent{\textbf{6. GIST}}: \\\\
\texttt{ \textcolor{red}{User: }The value for '\{target\_variable\}' is missing. Please provide a contextually relevant descriptor that can be used as a substitute for this missing value, based on scientific knowledge of CVD processes. Please provide only a few words (4 to 5) description for the descriptor with an example range that's not too wide, without any additional text.\\
\textbf{Note:} The GIST prompt is used to generate descriptors, after which the imputation is carried out by using GUIDE for this study. This could be combined with other prompts to create new imputation strategies if required.}

\subsection{Temperature or Entropy Selection}
For the ChatGPT-assisted imputation, we first use the \texttt{GPT-4o-mini} model for benchmarking. The suggested temperature for the \texttt{GPT-4o} model, which can be interpreted as the randomness or creativity of the model, is somewhere between 0.7 and 1, as mentioned in the \texttt{OpenAI}'s documentation. 

\begin{figure}[tp]
    \centering
    \includegraphics[width=0.7\textwidth]{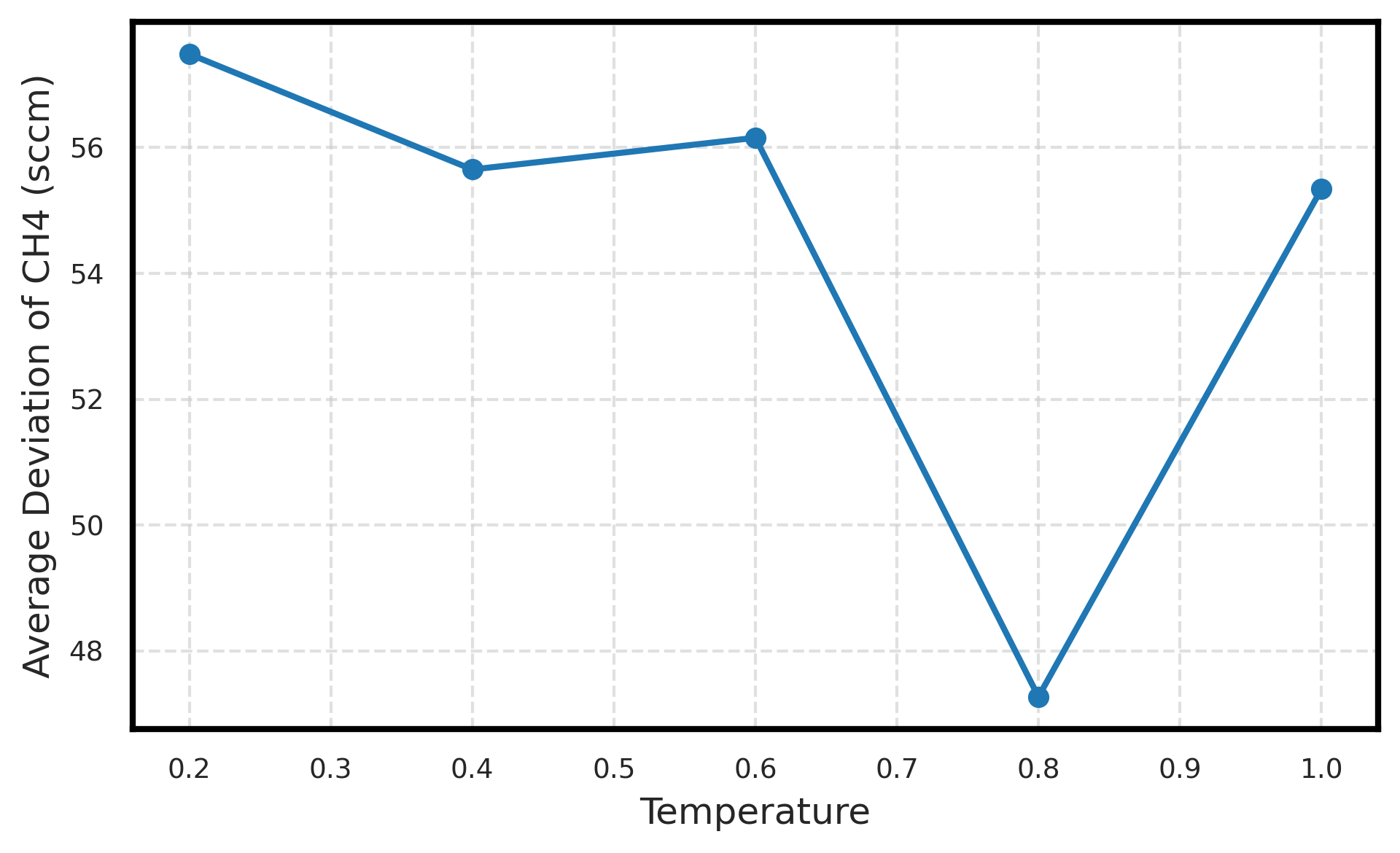}  
    \caption{Temperature versus imputation accuracy for \texttt{ChatGPT-4o-mini}.}
    \label{fig:temp}
\end{figure}

Benchmarking was performed by calculating the average deviation for \ce{CH4} imputation. This is done by first removing the ground truth and then imputing the rows for the artificially removed missing values using the GUIDE prompt. The difference between the imputed value and the previous ground truth is termed as the deviation. The deviations are averaged for the \ce{CH4} attribute over temperatures 0.2 to 1.0 in steps of 0.2. As shown in \ref*{fig:temp}, the lowest average deviation is achieved for a temperature of 0.8. For very low temperatures, we observe higher deviations, which show the lack of creativity of the model in adjusting its predictions over the six feedback iterations. Each iteration is meant to adjust the model's prediction near to ground truth, but a low temperature could prevent the model from learning through the feedback, making it deterministic. On the other hand, a very high temperature can make the model's prediction too random to follow any iterative adjustment. Temperature 0.8, on the other hand, enables some variation while not largely deviating from the true values of all other attributes.

\subsection{Benchmarking ChatGPT-4o-mini}
\begin{figure}[h!]
    \centering
    \includegraphics[width=\textwidth]{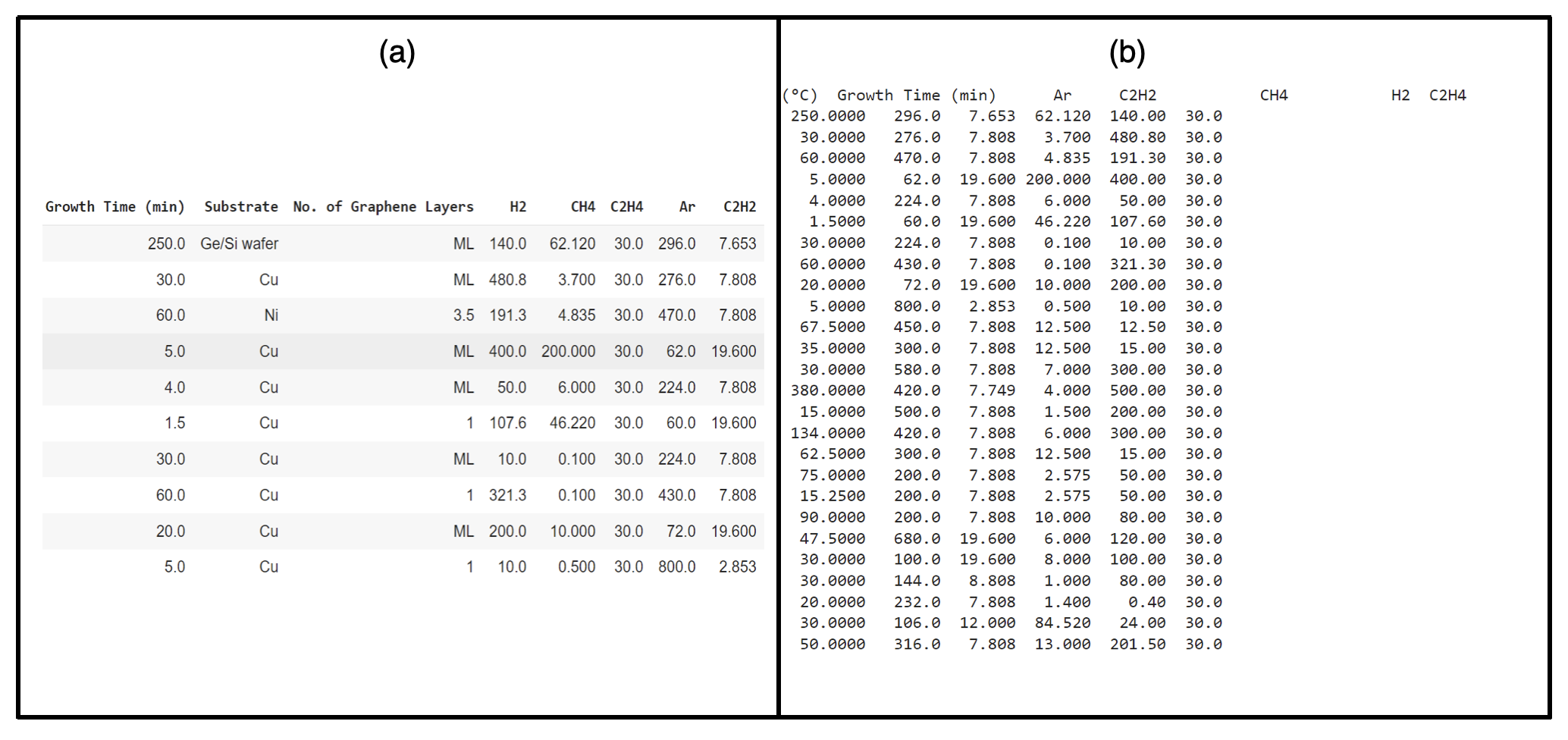}  
    \caption{Snippets from the output of the (a) KNN and (b) \texttt{ChatGPT-4o-mini} imputation.} 
    \label{fig:KNNPromptbenchmark}
\end{figure}

To use \texttt{ChatGPT} as an imputer, we first check if the language model can replicate the performance of a statistical imputer, like KNN. \texttt{ChatGPT} behaves exactly like a KNN if provided with the relevant prompt instruction. The following prompt is used to check for this capability:\\
\texttt{\textcolor{blue}{User Prompt: Please impute the missing values in the following dataset using statistical inference (like mean or median).}\\}
\ref*{fig:KNNPromptbenchmark} shows the KNN imputation and the text output by ChatGPT, clearly demonstrating that ChatGPT replicates the KNN imputation pattern using the
above instruction prompt. This can be directly observed with the imputed values of \ce{C2H4.} 

\begin{figure}[H]
    \centering
    \includegraphics[width=\textwidth]{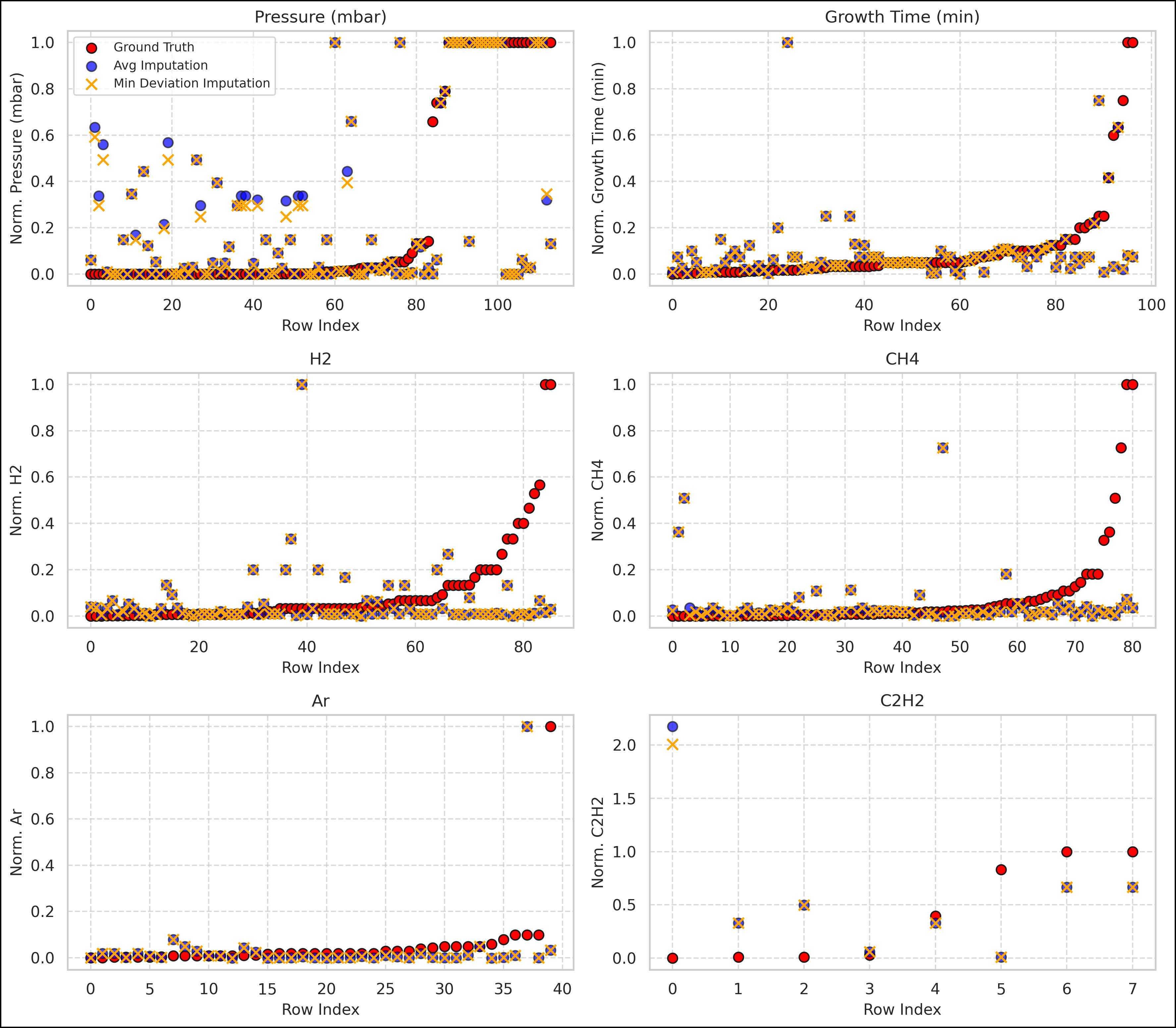}  
    \caption{Imputed values using the GIST prompt, either averaged over six iterations or taken as the minimum instance among the six iterations, for different normalized attributes, compared to the ground truth values.}
    \label{fig:raw_imputed_all}
\end{figure}

To evaluate the performance of each \texttt{ChatGPT} prompt imputer, we use the mean absolute error (MAE) as a metric, measuring the difference between imputed values and true values. The MAE is averaged over six iterations to account for any randomness in the imputation process.
The algorithm to perform the benchmarking analysis is given as follows:\\
\begin{enumerate}[label=\textbf{\arabic*}.]
    \item Extract rows where the target variable has known values (ground truth).
    \item Create a list to store predictions and deviations for these rows.
    \item For each ground truth row, construct a prompt using the experimental parameters as context.
    \item Use iterative predictions to impute the target variable value, refining based on deviations from the ground truth.
    \item Save the imputed predictions and their deviations for each ground truth row.
    \item Repeat this process for all target variables in the dataset.
\end{enumerate}
The analysis is carried out with normalized attributes to compensate for the scaling factor and different orders of magnitude of the attributes, as shown in \ref*{fig:raw_imputed_all}. We use a standard scaler to normalize the ground truth values between 0 and 1. Some imputation values fall above the normalized scale because they are plotted with respect to the ground truth normalized values. When scaling with the min-max scaler, we need to select the minimum and maximum values of either the ground truth or the imputed value range. This can lead to some of the imputed values falling out of the scale, which simply means that the imputed values are greater than the maximum ground-truth value for the corresponding attribute. The red, blue, and yellow points represent the ground truth, averaged imputes, and minimum imputed values, respectively. The averaged imputed values are computed over six iterations, providing an overall est. In contrast, the minimum imputed values correspond to the instance with the lowest deviation from the ground truth, highlighting the most accurate predictions. The imputed values for six attributes with the GIST prompt are shown in \ref*{fig:raw_imputed_all}. 

\subsection{Imputation Post Benchmarking}
The imputation process is straightforward after the benchmarking step and can be established with the following steps: 
\begin{enumerate}
    \item Identify rows in the dataset with missing values for the target variable.
    \item Set a retry limit to manage the number of imputation attempts (or iterations).
    \item For each row with a missing value, generate a prompt based on its contextual attributes.
    \item Invoke \texttt{ChatGPT-4o} to impute the missing value using the generated prompt.
    \item Replace the missing value in the dataset with the validated imputed value.
    \item Repeat the process for all rows with missing values until all are imputed or the retry limit is reached.
\end{enumerate}

\subsection{KNN Imputation}
We chose K=5 for the KNN imputation process, which accounts for a reasonable number of neighbors for the scarce dataset. Higher K values can bias the predictions, while lower K values might not capture the statistical variance. Moreover, $K=4$ and $K=6$ follow the same distribution as K=5 as shown in \ref*{tab:kvalue-effect} with the average MAE and MSE values (deviations from ground-truth) for the imputation of all attributes, allowing us to limit the value to $K=5$. 

\begin{table}[htbp]
  \centering
  \begin{tabular}{ccc}
    \hline
    \textbf{K Value} & \textbf{MAE} & \textbf{MSE} \\
    \hline
    4 & 0.2306 & 0.1147 \\
    5 & 0.2686 & 0.1252 \\
    6 & 0.2575 & 0.1613 \\
    \hline
  \end{tabular}
  \caption{Effect of K-value on imputation mean absolute error (MAE) and mean squared error (MSE).}
  \label{tab:kvalue-effect}
\end{table}

\begin{figure}[h!]
    \centering
    \includegraphics[width=\textwidth]{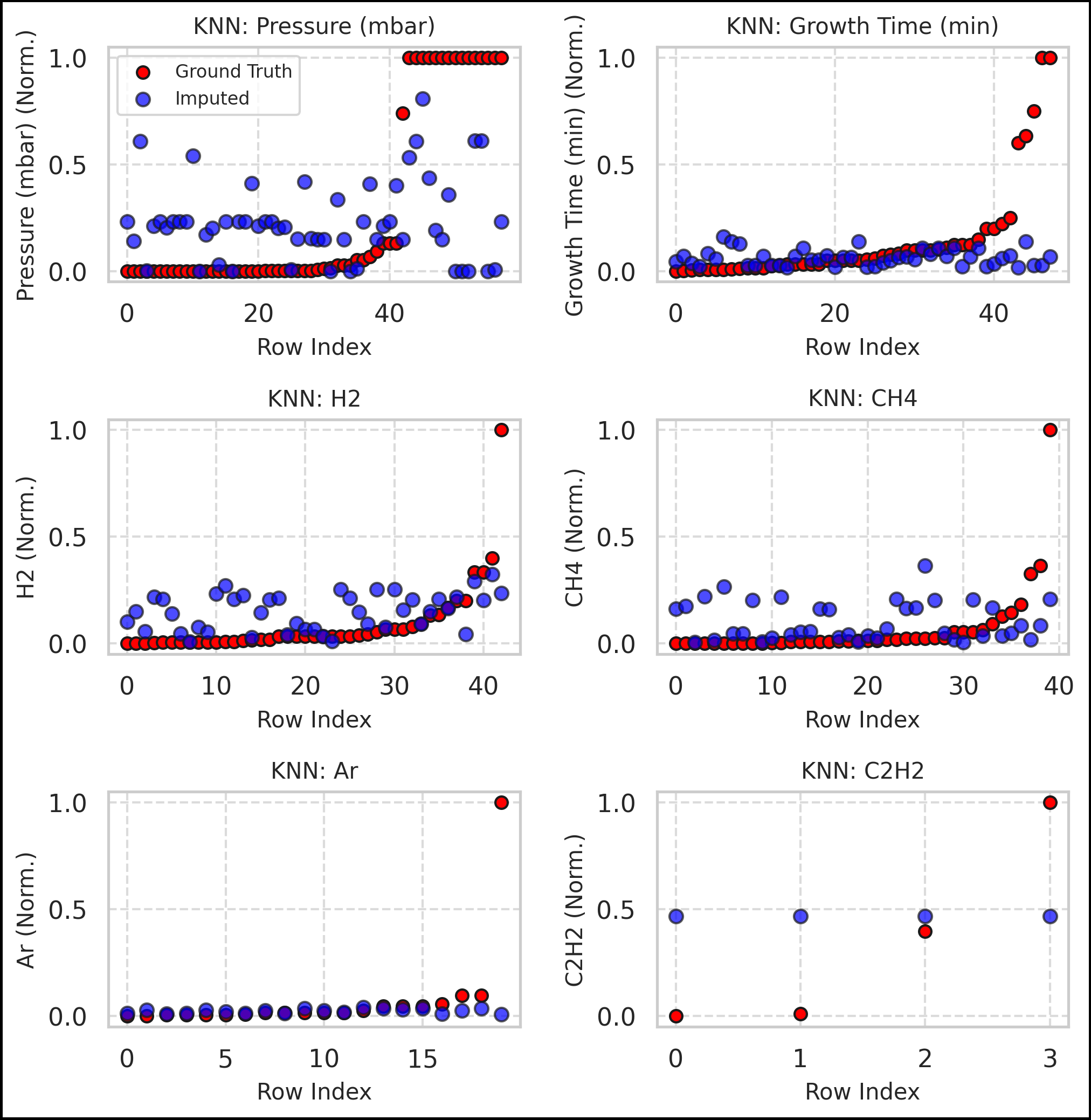}  
    \caption{Imputed values using KNN for different normalized attributes, compared to the ground truth values.}
    \label{fig:imputationknn}
\end{figure}

\ref*{fig:imputationknn} shows the imputation analysis for KNN. We use 50 \% of the ground truth data for KNN benchmarking while keeping the remaining 50 \% as a validation set. The \texttt{ChatGPT} imputation is carried on without any prior knowledge of the ground truth data, while the KNN requires some statistical distribution to work on. A higher split of training data would leave just a few points for validation, so we stick to 50 \% as a reasonable split. For the imputation step post benchmarking, we use all the known ground truth values. \ref*{fig:imputationknn} shows the KNN imputation benchmarking.

\section{Featurization}
The featurization of the substrate attribute involves addressing diverse categories and the absence of a standard format. Key techniques used include:
\begin{itemize}
    \item \textbf{One-hot encoding}: Not preferred due to the curse of dimensionality as it leads to too many features.
    \item \textbf{Label encoding}: Avoided for classification problems with many categories due to potential ordinal bias- categories might get related in terms of integral magnitude than labels.
    \item \textbf{Standardization using LLM \& label encoding}: Substrate descriptions are standardized using a ChatGPT-4o-mini to create fewer, physically meaningful categories, which can then be label encoded.
    \item \textbf{Text embedding using LLM}: Uniform vectors of desired dimensions are generated for different substrates using a model like \texttt{OpenAI}'s \texttt{text-embedding-3-small}, preserving semantic meaning.
\end{itemize}

\begin{figure}[h!]
    \centering
    \includegraphics[width=1.0\textwidth]{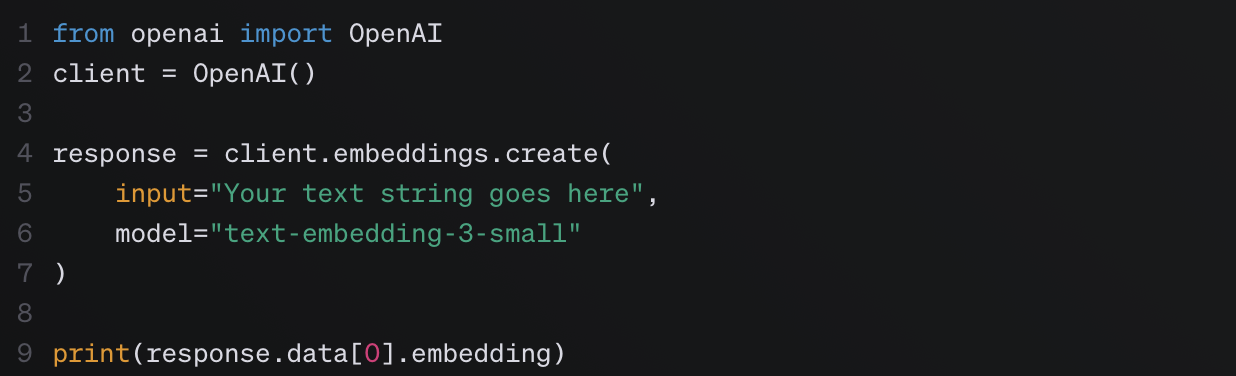}  
    \caption{\texttt{OpenAI} API call to create embedding vectors}
    \label{fig:api}
\end{figure}

For graphene layers, label encoding is used for binary, ternary, or quaternary classification, offering a compact representation of categories. \ref*{fig:api} shows the code snippet of how to interact with the \texttt{OpenAI} API to get the embeddings. 

\begin{figure}[h!]
    \centering
    \includegraphics[width=1.0\textwidth]{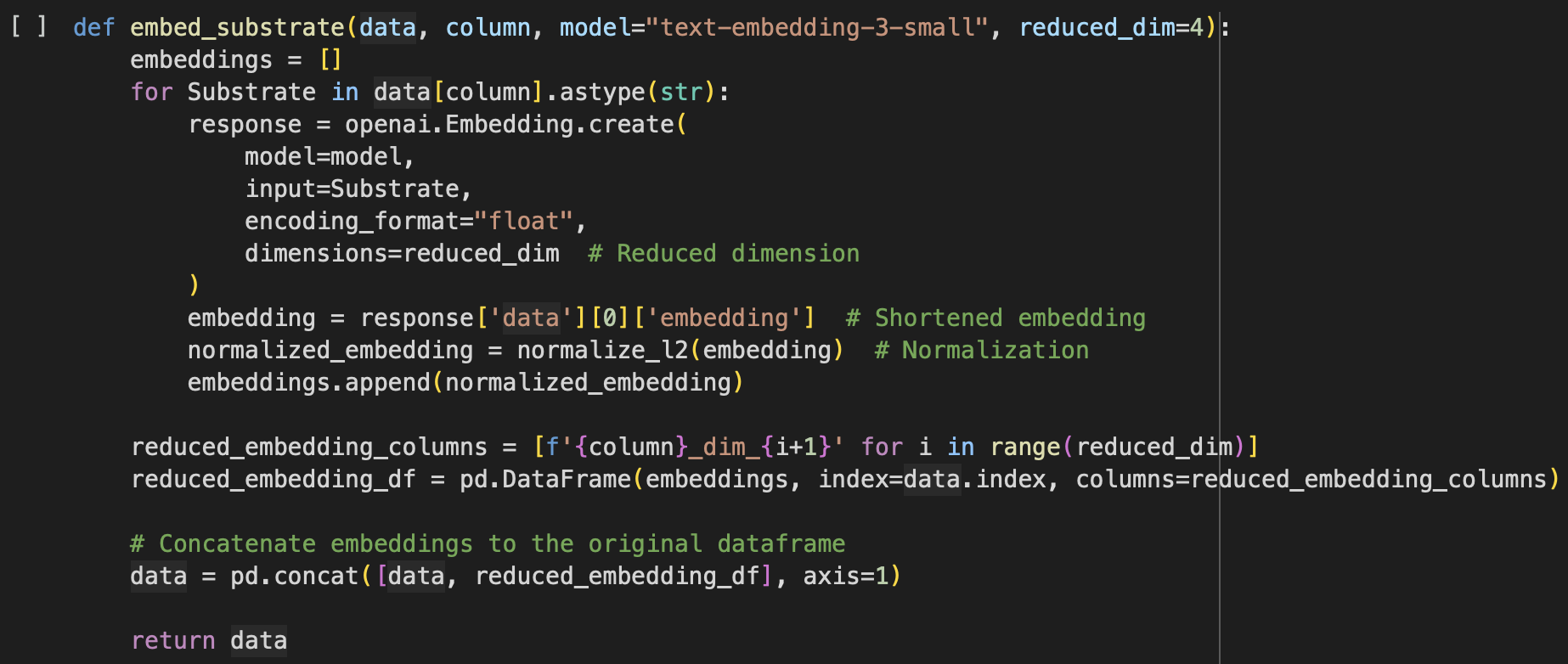}  
    \caption{Embedding featurization method with reduced dimensions.}
    \label{fig:embedding}
\end{figure}

\textbf{Note:} We use the continuous-valued attributes to test the featurization techniques and pick the best choice. It would be important to keep the featurization step different from the discretizing step, which would require its own testing before being combined with the featurization techniques.\\\\
Additionally, the prompt used to create standard descriptors using ChatGPT for the standardization using LLM and label encoding featurization method is given below.\\
\texttt{\textbf{Prompt: }\textcolor{blue}{
\textbf{Featurize the following substrate description: \texttt{\{substrate\}}.} \\
Return the output in a structured format using only a Python dictionary with these keys:
\begin{itemize}
    \item \texttt{'main\_material'}: Main material used in the substrate (e.g., 'Cu', 'Ni', 'SiO2').
    \item \texttt{'is\_alloy'}: 1 if it's an alloy, 0 otherwise.
    \item \texttt{'substrate\_type'}: Type of substrate (e.g., 'metal', 'semiconductor').
    \item \texttt{'surface\_coating'}: 1 if there's a surface coating, 0 otherwise.
    \item \texttt{'coating\_material'}: Coating material if available, 'None' if not.
    \item \texttt{'crystallographic\_orientation'}: Crystallographic orientation if available, 'Unknown' if not.
    \item \texttt{'is\_multilayer'}: 1 if it's a multilayer substrate, 0 otherwise.
\end{itemize}
Populate with the most probable values if some data is missing.
}}\\\\
An important step before training with the features is to reduce the dimensions of the embedding vectors to avoid the dimension problem. This can be manually done by principal component analysis, but we leverage \texttt{OpenAI}'s inbuilt method to reduce the dimensions to 4, 8, or 16 as these are within the dimension limit for the small dataset. \ref*{fig:embedding} shows the code snippet for the embedding method incorporated for this case.

Another step is to normalize the embeddings with \texttt{OpenAI}'s inbuilt l2 normalization method before concatenation with the original features. This is also highlighted in \ref*{fig:embedding}. To evaluate the impact of different featurization methods, we assess the performance of an SVM classifier on various featurized substrates. The tests are conducted using imputed data with the GUIDE prompt. \ref*{tab:substrate_features} shows a few examples of the LLM-standardized substrate features. These obtained features are processed through the \texttt{text-embedding-3-small} model to get the corresponding substrate embeddings.
\begin{table}[h!]
\centering
\begin{tabularx}{\textwidth}{|X|l|l|l|l|}
\hline
\textbf{LLM Standardized Text} & \textbf{Feature 1} & \textbf{Feature 2} & \textbf{Feature 3} & \textbf{Feature 4} \\
\hline
{\texttt{{'main\_material': 'Ge', 'is\_alloy': 0, 'substrate\_type': 'semiconductor', 'surface\_coating': 0, 'coating\_material': 'None', 'crystallographic\_orientation': 'Unknown', 'is\_multilayer': 0}}} & -0.1160 &-0.4696&0.3291&-0.1906\\ 
\hline
{\texttt{{'main\_material': 'Cu', 'is\_alloy': 0, 'substrate\_type': 'metal', 'surface\_coating': 0, 'coating\_material': 'None', 'crystallographic\_orientation': 'Unknown', 'is\_multilayer': 0}}} &0.3818&-0.4416& -0.2391& -0.0478 \\ 
\hline
{\texttt{{'main\_material': 'Ni', 'is\_alloy': 0, 'substrate\_type': 'metal', 'surface\_coating': 0, 'coating\_material': 'None', 'crystallographic\_orientation': 'Unknown', 'is\_multilayer': 0}}} &0.3459 &-0.1774 &-0.4724& 0.5758\\ 
\hline
\end{tabularx}
\caption{LLM-standardized substrate features, followed by embedding the standardized text using \texttt{text-embedding-3-small}.}\label{tab:substrate_features}
\end{table}

\begin{table}[htbp]
  \centering
  \begin{tabular}{|l|c|c|c|c|}
    \hline
    \textbf{Featurization Technique} & \textbf{Accuracy} & \textbf{Precision} & \textbf{Recall} & \textbf{F1} \\
    \hline
    Label Encoded Features 
      & 0.45 & 0.45 & 0.46 & 0.41 \\
    \hline
    LLM Proposed Substrate (Standardized) 
      & 0.51 & 0.51 & 0.52 & 0.51 \\
    \hline
    text-embedding-3-small (dim=4) 
      & 0.54 & 0.54 & 0.54 & 0.54 \\
    \hline
    text-embedding-3-small (dim=8) 
      & 0.57 & 0.58 & 0.58 & 0.57 \\
    \hline
    text-embedding-3-small (dim=16) 
      & 0.54 & 0.55 & 0.55 & 0.54 \\
    \hline
    Standardized text-embedding-3-small (dim=4) 
      & 0.58 & 0.58 & 0.58 & 0.57 \\
    \hline
    Standardized text-embedding-3-small (dim=8) 
      & 0.51 & 0.52 & 0.52 & 0.51 \\
    \hline
  \end{tabular}
  \caption{Classification performance metrics for different featurization methods.}
  \label{acc-report-featurization}
\end{table}

The featurization techniques are assessed by an SVM model, the details of which are given in \ref*{note:svm}. As shown in \ref*{acc-report-featurization}, the classification performance is highest for the substrate embedding method with eight reduced dimensions (SE-8), showing a substantial increase over the label encoded method.

\section{Discretization}
We discretize the variable space to address attributes with heavily skewed distributions, such as Pressure, Growth Time, and Gas Flow Rates. Training models over continuous values for these attributes often undermines accuracy and usability, as the model may fail to capture meaningful patterns in the data. Discretization simplifies the data and enhances model performance and interpretability by converting continuous values into categorical labels. In this study, we examine the following types of Discretization Techniques:
\begin{itemize}
    \item Equal-width binning
    \item Equal-frequency binning
    \item K-means binning
    \item Decision Tree binning
\end{itemize}

\begin{figure}[h!]
    \centering
    \includegraphics[width=1\textwidth]{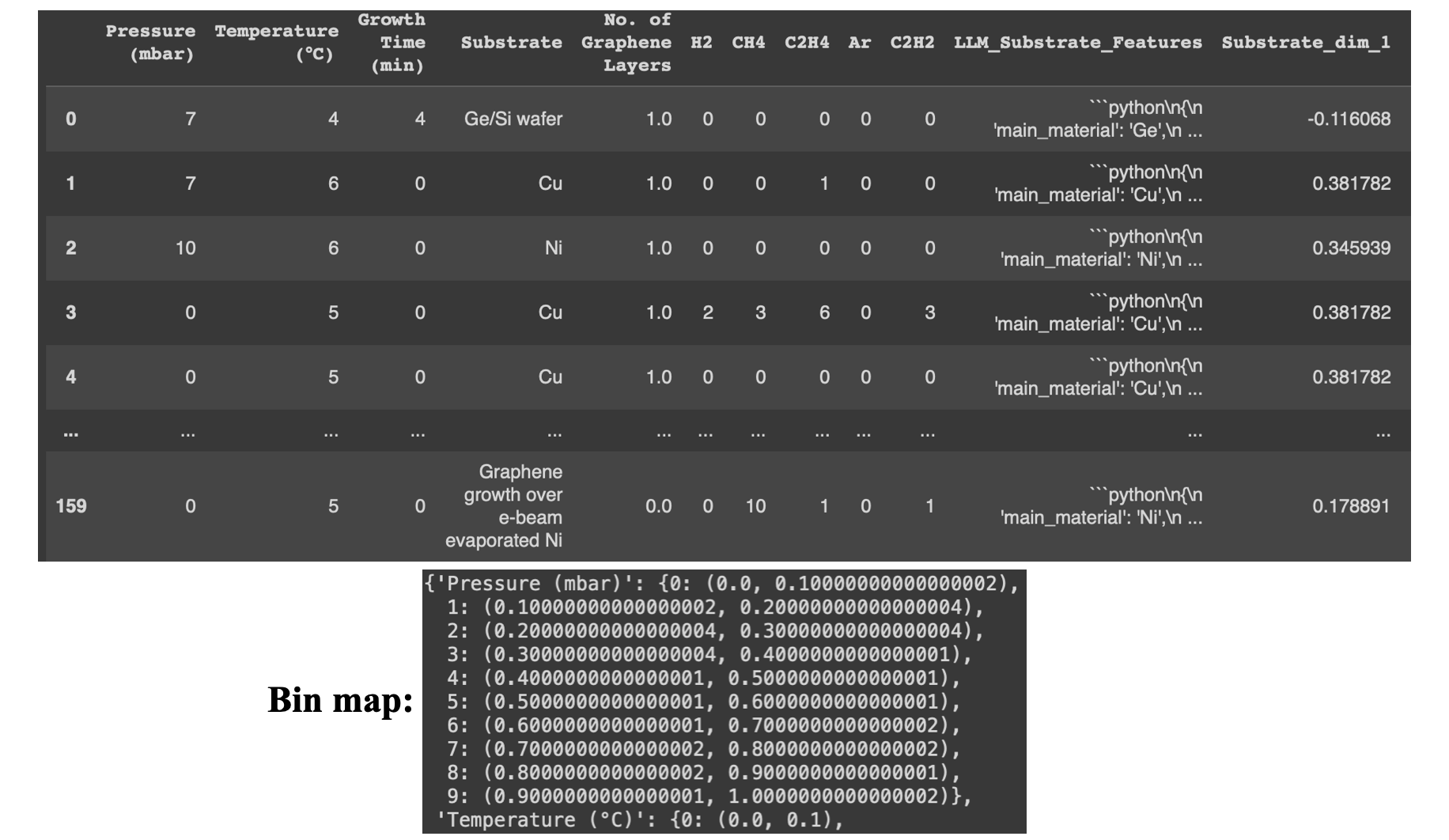}  
    \caption{An example dataset representation after integer encoding of discrete labels with equal width binning in 10 bins using standardized embeddings of substrates.}
    \label{fig:discretization}
\end{figure}

We use label encoding to discretize without increasing the dataset's dimensions. By assigning integer labels to the bins created through the selected discretization technique, we keep the transformed data compact while maintaining its interpretability.\\
\textbf{1. Equal-width Binning:}  
Divide the range into bins of equal width.  
\begin{equation}
w = \frac{\max(X) - \min(X)}{n}
\end{equation}
where \( w \) is the bin width, \( X \) represents the attribute values, \( n \) is the number of bins, and \( \min(X) \) and \( \max(X) \)  are the minimum and maximum values of attribute $X$, respectively. \\\\
\textbf{2. Equal-frequency Binning:}  
Each bin contains the same number of data points.  
\begin{equation}
f = \frac{N}{n}
\end{equation}
where \( f \) is the number of data points per bin, \( N \) is the total number of data points, and \( n \) is the number of bins.\\\\
\textbf{3. K-means Binning:}  
Groups similar values into \( k \) clusters  
\begin{equation}
J = \sum_{i=1}^{k} \sum_{x \in C_i} \|x - \mu_i\|^2
\end{equation}
where \( J \) is the within-cluster variance, \( C_i \) represents the \( i ^{\text{th}}\) cluster, \( \mu_i \) is the mean of the \( i^{\text{th}} \) cluster, and \( x \) is a data point.\\\\
\textbf{4. Decision Tree Binning:}  
Splits the data based on information gain calculated using the Gini impurity (or entropy) for classification, thereby minimizing the overall impurity in the child nodes. The information gained is defined in the following equation.
\begin{equation}
\Delta I = I_{\text{parent}} - \sum_{j} \frac{|C_j|}{|C|} I_j
\end{equation}
where \( \Delta I \) is the information gain, \( I_{\text{parent}} \) is the impurity of the parent node, \( C_j \) is the \( j ^{\text{th}}\) child node, \( |C_j| \) is the size of the \( j ^{\text{th}}\) child node, and \( I_j \) is the impurity of the \( j \)-th child node, and C is the total number of nodes.\\
The discretization techniques are validated using the SVM classifier with the same parameters used for the featurization testing. To keep track of the encoded labels used in the discretization step, we create a bin map that assigns each bin with its respective range, which can be retrieved later to uncover the meaning of those labels. \ref*{fig:discretization} shows a data frame with discretized variable space and its corresponding bin map. 
\begin{table}[htbp]
  \centering
  \begin{tabular}{|l|c|c|c|c|}
    \hline
    \textbf{Discretization Strategy} & \textbf{Accuracy} & \textbf{Precision} & \textbf{Recall} & \textbf{F1} \\
    \hline
    Equal-Width (10 bins) & 0.58 & 0.57 & 0.57 & 0.57 \\
    \hline
    Equal-Frequency (12 bins) & 0.55 & 0.54 & 0.54 & 0.55 \\
    \hline
    K-Means (K=6) & 0.51 & 0.52 & 0.52 & 0.51 \\
    \hline
    Decision-Tree (10 splits, 80-20) & 0.62 & 0.62 & 0.61 & 0.61 \\
    \hline
    Decision-Tree (10 splits, 90-10) & 0.65 & 0.64 & 0.64 & 0.65 \\
    \hline
    Decision-Tree Ternary (10 splits, 80-20) & 0.72 & 0.76 & 0.66 & 0.69 \\
    \hline
  \end{tabular}
  \caption{Classification performance metrics for different discretization methods applied on GUIDE-SE(4) method.}
  \label{tab:binning-accuracy}
\end{table}

\ref*{tab:binning-accuracy} shows the performance of the SVM classifier on the equal-width binning technique with 10 bins as an example. The dataset is the GUIDE-imputed SE(4)-featurized discrete space used to classify the graphene layers into single layer versus multi-layer. The performance on attributes discretized by equal-frequency binning is shown is also shown with 12 bins as an example.
For the case of K-means binning, the data is reshaped into a single column, and clustering is performed iteratively for \( K \) values ranging from 2 to a maximum 10 or dataset size. The optimal number of clusters is determined using the elbow method by minimizing the change in distortions. Finally, the data is assigned cluster labels corresponding to the optimal \( K \). \ref*{fig:K-mean-code} shows the code snippet for the implementation of the K-means binning technique. The final discretization technique involves the use of decision trees, which create labels for the attributes based on the target value (i.e., the number of graphene layers) in a supervised manner. 

\begin{figure}[H]
    \centering
    \includegraphics[width=0.8\textwidth]{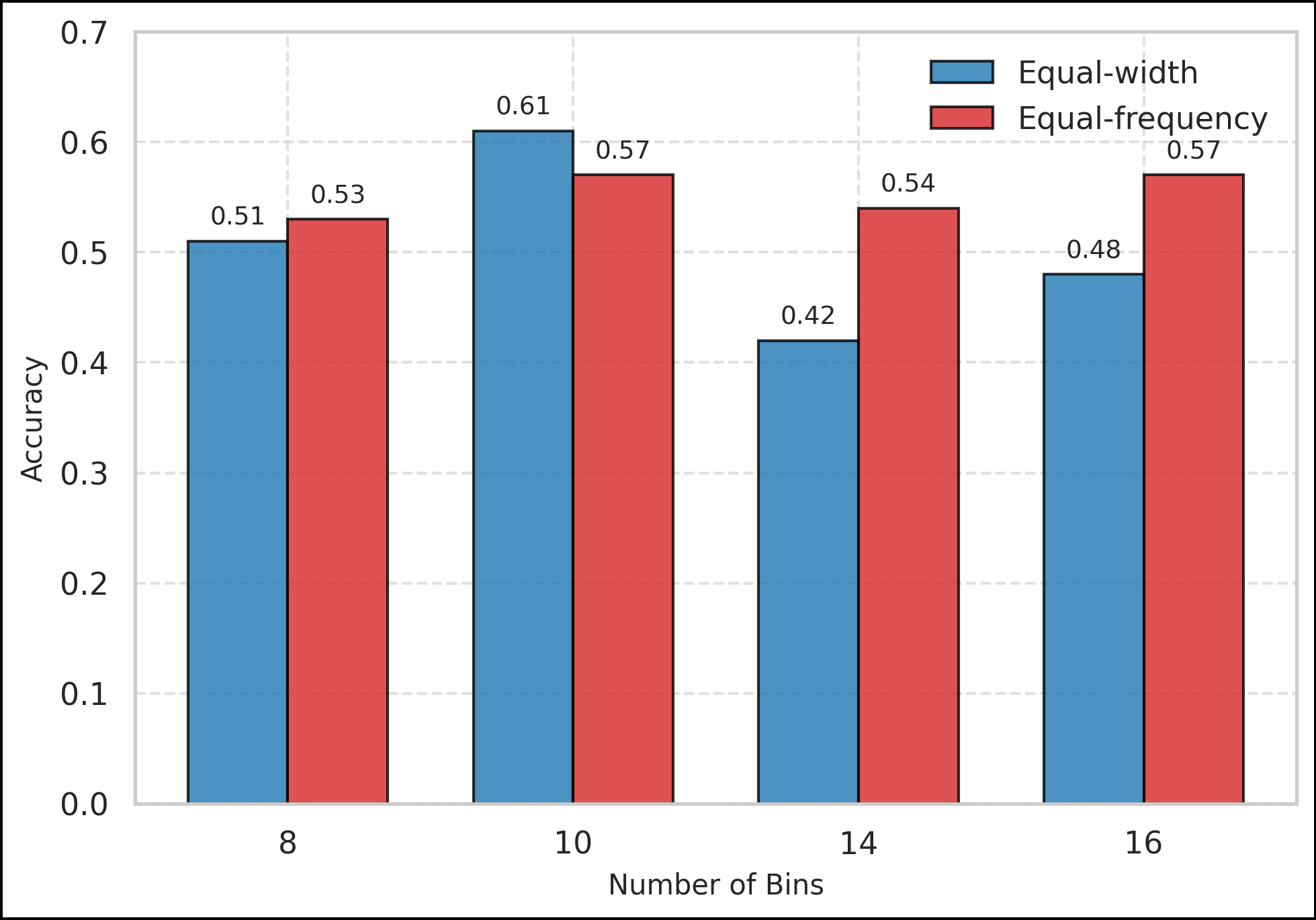}  
    \caption{Effect of the number of bins in equal-width binning on the accuracy of the SVM classifier.}
    \label{fig:svm_numbin}
\end{figure}

\begin{figure}[H]
    \centering
    \includegraphics[width=1\textwidth]{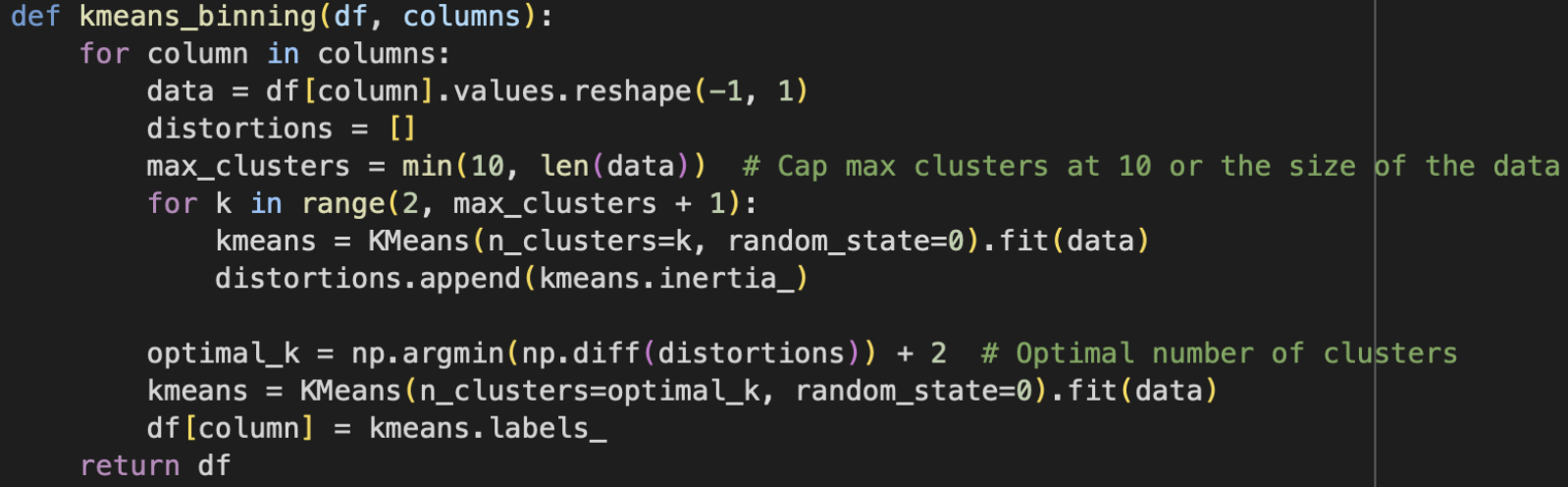}  
    \caption{Function for K-means binning with the selection of an optimal number of bins.}
    \label{fig:K-mean-code}
\end{figure}

The effect of the number of bins on the accuracy of equal-width and equal-frequency binning techniques is shown in \ref*{fig:svm_numbin}.

\section{Additional Machine Learning Details}
The choice of the ML architecture is primarily related to the learning capacity of the model. In the main manuscript, we present the learning rates of different ML models and conclude that the support vector machine (SVM) classifier is the optimal choice for the current dataset. 
For the binary classification problem, our dataset contains 83 monolayer and 81 instances of multilayer graphene. The multilayer category can be further divided into 31 bilayer and 50 multilayer instances for the ternary classification. The number of layers is label-encoded as follows: 0 for monolayer, 1 for bilayer, and 2 for multilayer (for ternary classification). The data are split into an 80-20 train-test ratio, with extensions to a 90-10 split for incremental learning curve analysis. 
\subsection{Learning Curve}
\begin{figure}[h!]
    \centering
    \includegraphics[width=1\textwidth]{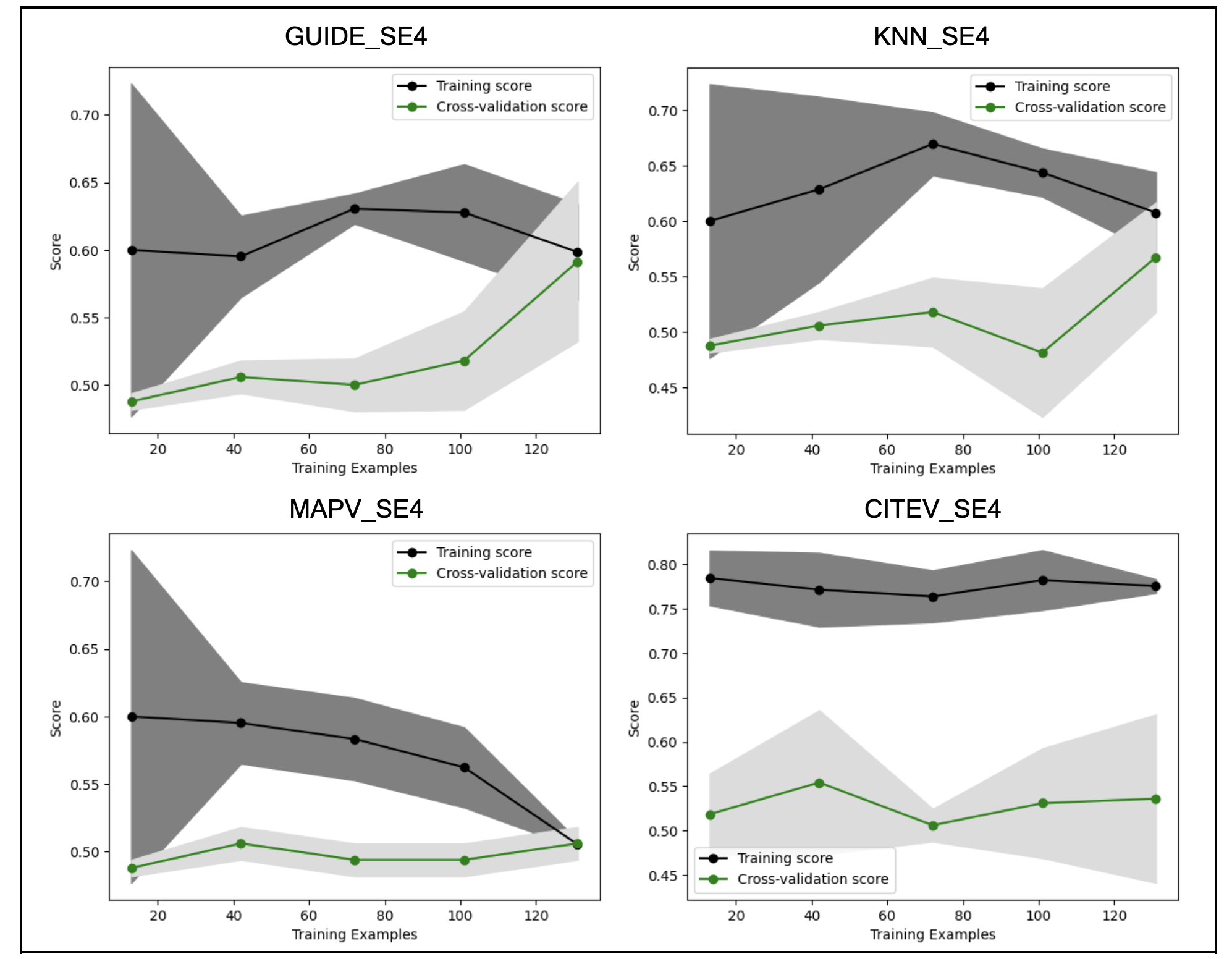}  
    \caption{Binary SVM classifier learning curves for different prompt imputed datasets with SE(4) featurization.}
    \label{fig:learning_curves}
\end{figure}

\begin{figure}[h!]
    \centering
    \includegraphics[width=1\textwidth]{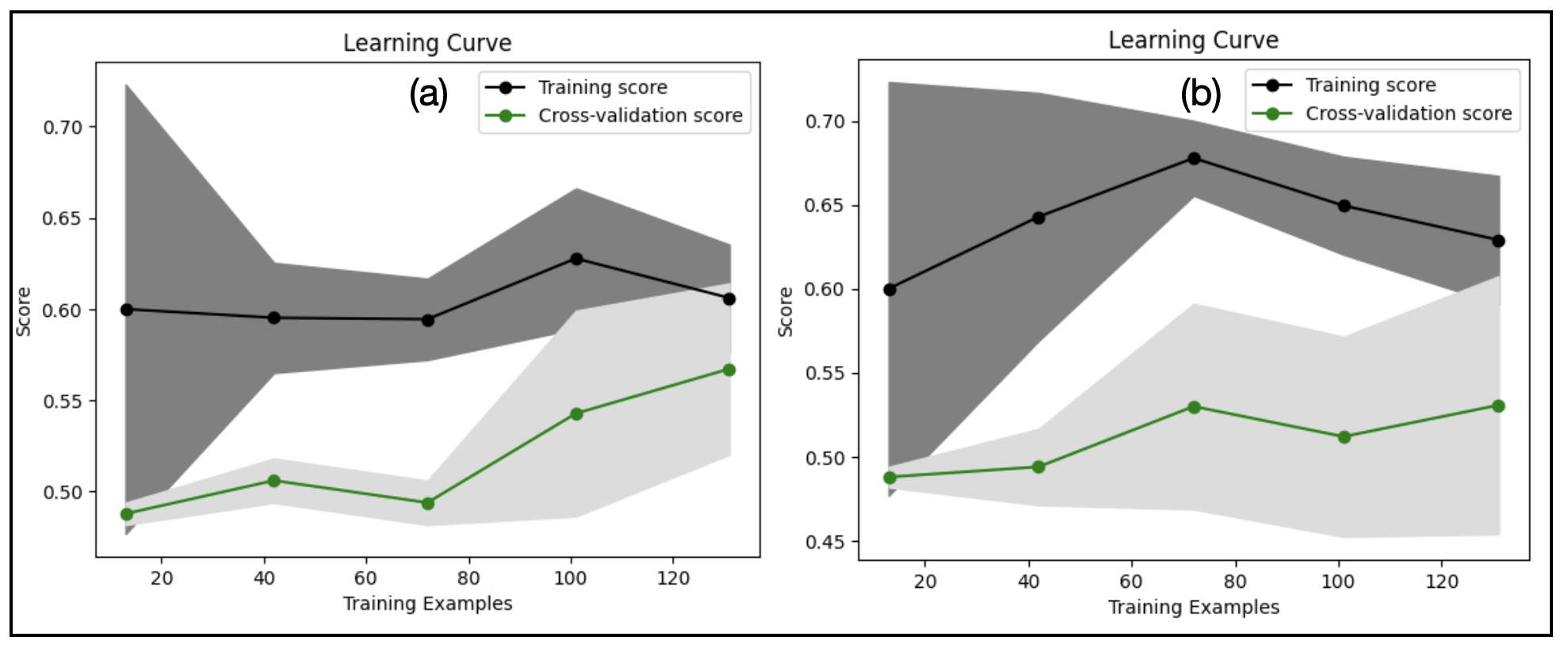}  
    \caption{Binary SVM classifier learning curves for prompts (a) MAP (b) CITE.}
    \label{fig:map_cite_learning}
\end{figure}

We use the \texttt{learning\_curve} function from the \texttt{scikit-learn} library with training sizes obtained via \texttt{np.linspace (0.1,1,5)} with 5-fold cross-validation (cv=5) to compute the learning curve. This technique involves splitting the dataset into five subsets of training sizes and five folds of validation sets, where the model is trained on each training subset and tested on each corresponding validation set. The training scores are obtained by evaluating the model on different training data sets for each fold. The test score is obtained by evaluating the model on the respective validation fold. Averaging the training and test scores, we get the learning and cross-validation curves, respectively. A clear description of this method can be found in the scikit-learn library or the Supplementary file Python notebook.
The binary SVM classifier learning curves for different prompt-based predictions using the SE-4 featurization technique are shown in \ref*{fig:learning_curves}. Additionally, the same techniques are applied to the datasets imputed by the MAP and CITE prompts, and their learning curves are obtained using the SVM classifier as shown in \ref*{fig:map_cite_learning}.
\subsection{SVM Model Architecture}\label{note:svm}
The decision boundary (hyperplane) is given by
   $ \mathbf{w} \cdot \mathbf{x} + b = 0$
where \( \mathbf{w} \) is the weight vector normal to the hyperplane, \( \mathbf{x} \) is the input vector, and \( b \) is the bias term.
During training, SVM optimizes the following objective function:
\begin{equation}
   \min_{\mathbf{w}, b, \boldsymbol{\xi}} \frac{1}{2} \|\mathbf{w}\|^2 + C \sum_{i=1}^n \xi_i, \quad 
\text{subject to: } y_i (\mathbf{w} \cdot \mathbf{x}_i + b) \geq 1 - \xi_i, \, \xi_i \geq 0. 
\end{equation}
where \( \xi_i \) is the slack variable for the \( i ^{\text{th}}\) data point, \( C \) is the regularization parameter, \( y_i \) is the class label (e.g, \(+1\) or \(-1\) in a binary classification), and \( \mathbf{x}_i \) is the \( i ^{\text{th}}\) data point.
A dual formulation enables the use of the kernel trick:
\begin{align}
    \max_{\boldsymbol{\alpha}} \sum_{i=1}^n \alpha_i - \frac{1}{2} \sum_{i=1}^n \sum_{j=1}^n \alpha_i \alpha_j y_i y_j K(\mathbf{x}_i, \mathbf{x}_j),  \\
    \text{subject to: } \sum_{i=1}^n \alpha_i y_i = 0, \, 0 \leq \alpha_i \leq C.
\end{align}
where \( \alpha_i \) is the Lagrangian multiplier for the \( i ^{\text{th}}\) data point, and \( K(\mathbf{x}_i, \mathbf{x}_j) \) is the kernel function that maps input data into a higher-dimensional space.
The regularization parameter (C) plays a crucial role in adjusting the trade-off between maximizing the margin and minimizing classification errors, particularly in scarce data with biased attributes. While custom regularization techniques are beyond the scope of this study, we recommend exploring such approaches as a potential strategy to enhance the accuracy of an SVM classifier.
\subsection{Stratification}
 This is done by using \texttt{stratify=y\_test} in the \texttt{train\_test\_split} function. Not using a stratified split could result in higher and falsified accuracy values if only a few samples of either positive or negative classes are present in the validation set.
\subsection{Hyperparameter Optimization}
We optimize the SVM hyperparameters by exhaustively searching over a predefined parameter grid for the regularization parameter $C$ (0.1, 1, 10, 100), the kernel coefficient $\gamma$ (\texttt{``scale''}, \texttt{``auto''}, 0.01, 0.1, 1, 10), and the kernel type (RBF, linear, polynomial). This search uses \texttt{GridSearchCV} with 5-fold cross-validation, where each hyperparameter combination is evaluated by its average accuracy across the folds. The best-performing set of parameters is then automatically selected, and the final SVM model is retrained on the entire training set before being tested on the held-out data.

\section{Fine Tuning \texttt{ChatGPT}}
\subsection{Dataset Preparation}
\texttt{OpenAI} specifies the requirement of a JSONL format dataset consisting of a System, User, and Assistant prompt, with the first one requiring general instructions, the second one corresponding to the input prompt, and the final one being the output or response desired from the model.
We write a simple function that specifies general classifier instruction for CVD and the name of the attributes, the user prompt consisting of the actual values of these attributes, and the assistant prompt specifying integers- 0,1, or 2- corresponding to monolayer, bilayer, or multilayer graphene, respectively. We divide the dataset into training and validation splits (80-20) beforehand and specify them during the job creation. The datasets and data preparation scripts are provided on our GitHub page.

\subsection{Job Creation}
We submit our training and validation datasets to the \texttt{OpenAI} job module, along with the specification of (a) the number of epochs to 3, (b) the learning rate multiplier to 1.0, (c) the batch size to 1, and (d) the model suffix (e.g., 4o or 4o-mini). The fine-tuning job terminates upon full validation over the three epochs, and the results are obtained as outputs of the job.

\subsection{Fine Tuning Results}
\begin{figure}[h!]
\includegraphics[width=1\textwidth]{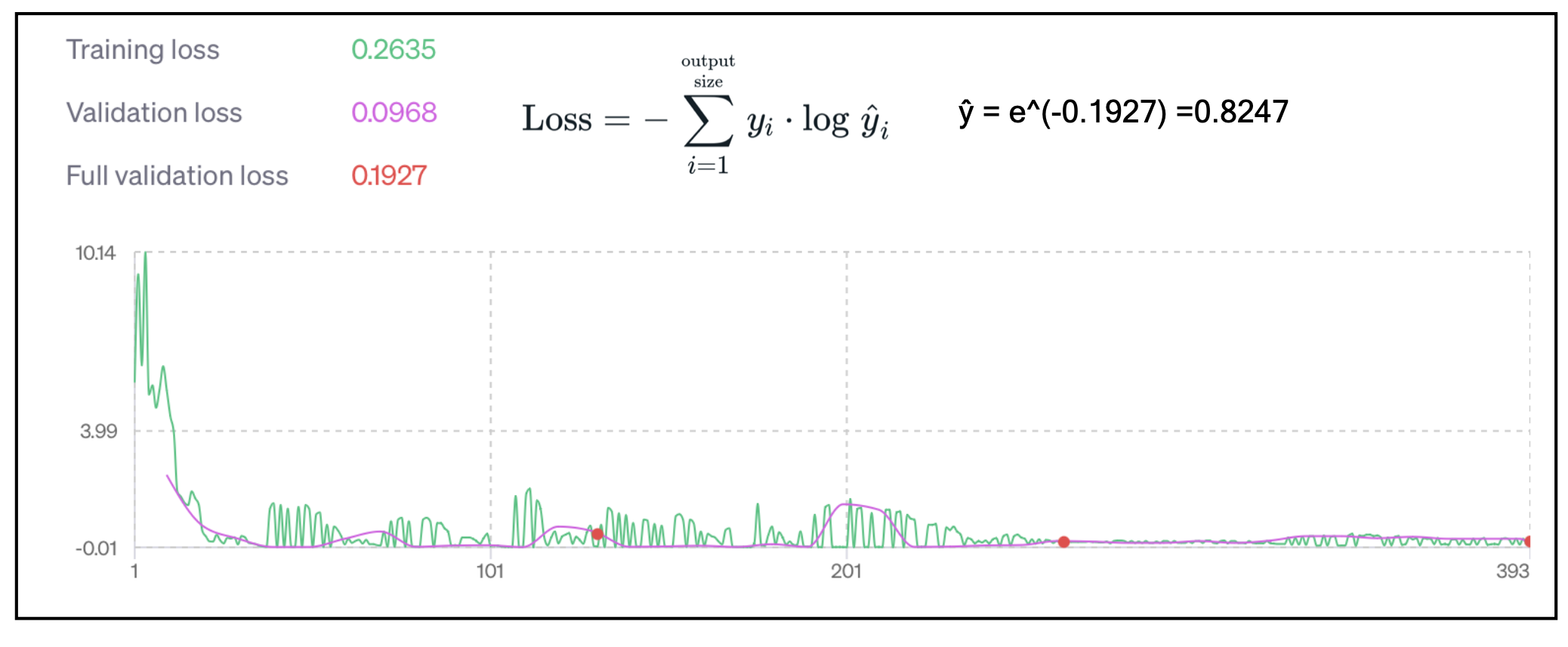} 
    \caption{Cross-validation curve obtained from \texttt{OpenAI} job completion output of \texttt{ChatGPT-4o}. The red dots on the curve indicate the completion of each epoch. The token prediction score, $\hat{y}$, is computed from the cross-entropy loss for the GUIDE imputed data.}
    \label{fig:fine-tuning-validation}
\end{figure}

The fine-tuning performance is evaluated by analyzing the cross-validation curves obtained after the job completion. The full validation score represents the cross-entropy loss that can be translated to the token prediction accuracy as shown in \ref*{fig:fine-tuning-validation} which is for a fine-tuned \texttt{ChatGPT-4o} model.
The token prediction accuracy does not represent the graphene layer classification accuracy but only the accuracy with which the output embeddings are predicted by  \texttt{ChatGPT}. We write a function to get the classification results on the validation set. The classification results are then used to calculate performance metrics, including the accuracy, precision, recall, and F1 score. \ref*{tab:acc-comparison} shows the results of different fine-tuned models with the cross-entropy accuracy and the actual classification accuracies.

\begin{table}[htbp]
  \centering
  \begin{tabularx}{\textwidth}{X c c c c c}
    \hline
    \textbf{Model} & \textbf{Token Accuracy} 
      & \textbf{Accuracy} & \textbf{Precision} & \textbf{Recall} & \textbf{F1} \\
    \hline
    Binary mini + GUIDE-SE(4) 
      & 0.7596 & 0.5152 & 0.5238 & 0.6471 & 0.5789 \\\\
    Ternary mini + GUIDE-SE(4) 
      & 0.6414 & 0.4545 & 0.4489 & 0.3676 & 0.3714 \\\\
    Binary featurized mini + GIST-SE(8)-DT
      & 0.4879 & 0.4545 & 0.4000 & 0.4000 & 0.4000 \\\\
    Ternary featurized mini + GUIDE-SE(4)-DT
      & 0.7116 & 0.4545 & 0.3667 & 0.3608 & 0.2803 \\\\
    Binary 4o + GUIDE-SE(4)
      & 0.8247 & 0.4545 & 0.4286 & 0.1765 & 0.2500 \\\\
    Ternary featurized 4o + GUIDE-SE(4)-DT
      & 0.7655 & 0.4888 & 0.4839 & 0.4444 & 0.3741 \\\\
    Binary 4o + KNN
      & 0.7525 & 0.4848 & 0.8000 & 0.2000 & 0.3200 \\\\
    Ternary 4o + KNN
      & 0.5265 & 0.5152 & 0.5000 & 0.3667 & 0.2828 \\\\
    \hline
  \end{tabularx}
  \caption{Comparison of cross-entropy accuracy for token prediction versus graphene layer classification accuracy, precision, recall, and F1 score.}
  \label{tab:acc-comparison}
\end{table}


\end{document}